\documentclass[preprint]{aastex}
\usepackage{psfig}

\begin{document}

\slugcomment{}

\title{Stellar Populations of Dwarf Elliptical Galaxies: UBVRI Photometry of Dwarf Elliptical Galaxies in the Virgo Cluster
\footnote{Based on observations with the VATT: the Alice P. Lennon Telescope and the Thomas J. Bannan Astrophysics
Facility.} 
}

\author{Liese van Zee}
\affil{Astronomy Department, Indiana University, 727 E 3rd St., Bloomington, IN 47405}
\email{vanzee@astro.indiana.edu}
\author{Elizabeth J. Barton\footnote{Hubble Fellow}}
\affil{Steward Observatory, University of Arizona, 933 N Cherry St., Tucson, AZ 85721}
\email{ebarton@as.arizona.edu}
\and
\author{Evan D. Skillman}
\affil{Astronomy Department, University of Minnesota, 116 Church St. SE,
Minneapolis, MN 55455}
\email{skillman@astro.umn.edu}

\begin{abstract}
We present UBVRI surface photometry for 16 dwarf elliptical galaxies in the
Virgo Cluster with previously measured kinematic properties.
The global optical colors are red, with median values for
the sample of 0.24 $\pm$ 0.03 in (U--B), 0.77 $\pm$ 0.02 in (B--V),
and 1.02 $\pm$ 0.03 in (V--I).
We recover the well known color-magnitude relation for cluster galaxies,
but find no significant difference in dominant stellar
population between rotating and non-rotating dwarf elliptical galaxies;
the average age of the dominant stellar population is 5-7 Gyr in
all 16 galaxies in this sample.  
Analysis of optical spectra confirm these age estimates and
indicate Fe and Mg abundances in the range of 1/20th to 1/3 of solar,
as expected for low luminosity galaxies.  Based on 
Lick indices and simple stellar population models, the
derived [$\alpha$/Fe] ratios are sub-solar to solar,
indicating a more gradual chemical enrichment history for dEs
as compared to giant elliptical galaxies in the Virgo Cluster.  These observations confirm
the marked difference in stellar population and stellar distribution
between dwarf and giant elliptical galaxies and further substantiate
the need for alternative evolutionary scenarios for the lowest mass cluster
galaxies.  We argue that it is likely that several different physical mechanisms
played a significant role in the production of the Virgo cluster dE galaxies
including {\sl in situ} formation, infall of dEs that were once part of
Local Group analogs, and transformation of dwarf irregular galaxies by
the cluster environment.  The observations support the hypothesis
that a large fraction of the Virgo cluster dEs are formed by ram pressure 
stripping of gas from infalling dIs.
\end{abstract}


\section{Introduction}

Dwarf elliptical galaxies are ubiquitous in high density regions.
For example, in the Local Group, dwarf elliptical galaxies out
number high luminosity galaxies by a factor of 6 \citep{M98},
and more than 50\% of the galaxies in the Virgo cluster are dEs
\citep{SBT85}.  However, while the clustering properties of dwarf 
elliptical galaxies appear to be an extension of the well known 
morphology-density relation for giant elliptical galaxies \citep[e.g.,][]{D80,PG84},
their structural parameters suggest that dwarf ellipticals are not 
merely low mass versions of elliptical galaxies \citep[e.g.,][]{LF83,K85,IWO86,CB87,BC91}. 
Unlike their giant counterparts, gas-poor dwarf elliptical
galaxies and gas-rich dwarf irregular galaxies have similar
stellar distributions which suggests that passive evolution of the stellar population 
could result in morphological evolution
from one dwarf classification to the other \citep[e.g.,][]{LF83,K85,DP88}.
In fact, many of the key morphological 
differences between dwarf elliptical and dwarf irregular galaxies 
can be traced to the cessation of star formation activity rather than
to dynamical differences between the two classes. 
In support of this, \citet{STCDSGDM03} note that the star formation histories of
dwarf elliptical galaxies may be very similar to those of dwarf irregular galaxies
right up to the point of the loss of their gas.  
These similarities
led to the hypothesis that dwarf elliptical galaxies are evolved versions
of dIs whose ISM has been disrupted by either internal \citep[e.g., blow-out,][]{DS86}
or external \citep[e.g., ram pressure stripping,][]{LF83} processes.
In addition to similar stellar distributions, recent kinematic studies of dEs in 
the Virgo cluster indicate that a large fraction of dEs and dS0s
have a significant rotational component \citep{PGCSBG02,GGvM03,vSH04}.
Furthermore, the amplitudes of the dE rotation curves are comparable to those
of similar luminosity dwarf irregular galaxies \citep{vSH04}. 
These results support the possibility that some cluster dEs are
evolved dwarf irregular galaxies which have lost their ISM and thus
have little-to-no star formation activity at the present epoch.

The possibility that some dIs evolve into dEs 
is almost tautological since dEs must have been gas-rich, star forming
low mass galaxies at some point in the past.  Thus, the key question for
dwarf galaxy evolution is not {\it whether} dEs evolve from dIs, but, 
rather, {\it when} and {\it why} do some dwarf irregular galaxies lose
their ISM and stop forming stars.  The strong clustering properties
of dEs \citep[e.g.,][]{BTS90} suggest that environmental influences may play a
key role in the gas-loss process.  Possible evolution mechanisms include triggered
bursts of star formation with subsequent blow-out of the ISM \citep[e.g.,][]{DS86},
or ram pressure stripping of the ISM as the galaxy plunges through
the hot intra-cluster gas \citep[e.g.,][]{LF83}.  Both of these scenarios will 
result in an overabundance of dwarf elliptical galaxies in high density regions,
but the resulting star formation histories are quite different.  In a triggered
starburst scenario, most of the star formation activity will take place in
discrete events which may be correlated with passage through the cluster.
On the other hand, if the ISM is lost through ram pressure stripping,
most of the star formation activity will take place prior to the passage
through the cluster, and may occur in a more quiescent manner, typical
of dIs \citep{vZ01}.  Thus, examination of the stellar populations and
relative chemical enrichments in dwarf elliptical galaxies may provide 
insight into which of these mechanisms is the dominant process for the 
evolution of dIs to dEs in high density regions.

Within the Local Supercluster, the Virgo cluster is the nearest
high density region and thus provides an excellent environment
in which to explore stellar populations and kinematic properties
of low luminosity galaxies.
At the same time, Virgo is dynamically young \citep[e.g.,][]{TS84}
and contains several substructural units \citep{BTS87}; thus,
with a well selected sample it may be possible to study the 
evolution of dwarf elliptical galaxies within the cluster itself.
Recent observations of surface brightness fluctuations \citep{JBB04} have shown that the
dE population is stretched in depth similarly to the bright elliptical population
\citep{NT00,WB00} but not as extended as claimed previously by \citet{YC95}.
Furthermore, nucleated dEs appear to be more strongly clustered than
non-nucleated dEs in the cluster \citep[e.g.,][]{BTS87}.
In this paper, we examine broad band optical colors 
and optical spectroscopy of 16 dwarf elliptical galaxies in the Virgo Cluster
to investigate their star formation histories and to search for
correlations between kinematic properties and past star formation
activity.  The sixteen galaxies in this sample were selected as part of
a kinematic study of dwarf elliptical galaxies in Virgo \citep{vSH04};
approximately half of the sample have a significant rotational component.

As discussed in \citet{vSH04}, it is likely that the
rotating dwarf elliptical galaxies are remnants of dIs that
have been stripped of their gas;  the rotating dwarf elliptical galaxies
tend to be located near the outskirts of the cluster and their rotation 
curves and maximum amplitude velocities are comparable to similar
luminosity dwarf irregular galaxies.  At the same time, the
origin of the non-rotating dwarf ellipticals is unclear.  
Are these low luminosity galaxies which began with little-to-no angular
momentum?  Were they formed through the collision
of even lower mass galaxies and thus ended up with no net angular
momentum?  Did they transfer their angular momentum as they
moved through the dense cluster?  In support of the latter process,
 \citet{Me01a,Me01b} have shown that ``tidal stirring'' --
or repeated tidal shocks suffered by dwarf satellite galaxies -- can 
remove the kinematic signature from a low mass galaxy without 
significant alteration of its stellar distribution.  While it is 
difficult to distinguish between these scenarios based on the data at hand,  
the common feature of all of these scenarios is that non-rotating dwarf 
elliptical galaxies may have been significantly altered by their interaction 
with the cluster environment.  Thus, it is possible that the galaxies which still
retain their kinematic signature have been accreted relatively
recently while the non-rotating dwarf elliptical galaxy sample 
may contain galaxies with a wide range of cluster accretion times,
from the very recent to several Gyr in the past.
Since it is likely that the age of the dominant stellar population 
is indicative of the time of accretion,
observations of the stellar populations in these galaxies
could provide insight into the accretion timescales.

In this paper, we analyze the past star formation activity
in Virgo dwarf elliptical galaxies based on the combination
of stellar population synthesis and spectral synthesis models.
The imaging and spectroscopic observations are described in 
Section 2.  We discuss the derived optical structural parameters
of the dwarf elliptical galaxies in Section 3.  In
Section 4, we analyze the optical colors in the context
of simple stellar populations.  In Section 5, 
optical spectra are analyzed in the context of simple stellar population 
models to investigate typical timescales for chemical enrichment.
The implications for the observed stellar population ages and
enrichment histories are discussed in Section 6.
The conclusions of the paper are summarized in Section 7.

\section{Observations}

Optical images were obtained of 16 dwarf elliptical galaxies 
with the VATT 1.8m and the WIYN 3.5m telescopes.  
Optical spectra were obtained with the Palomar 5m telescope.
The kinematic results of the spectroscopic observations are presented in \citet{vSH04};
here, we analyze the spectra in terms of stellar population models.  The sample
selection and observations are described in this section.

\subsection{Sample Selection}
As described in \citet{vSH04},
the sample was selected to optimize the detection of rotational motion 
in dE galaxies.
Sixteen dwarf elliptical galaxies were selected from the Virgo Cluster Catalog
\citep[VCC,][]{VCC} based on their apparent luminosity (m$_B < 15.5$),
morphological classification (dwarf elliptical or dS0), and apparent
ellipticity ($\epsilon > 0.25$).  The latter criterion was imposed
to maximize the projected rotational velocity in the kinematic observations.
In addition, the sample was restricted to those galaxies with
measured recessional velocities that place them within the Virgo cluster,
at an approximate distance of 16.1 Mpc \citep[e.g.,][]{Ke00}.
Further details of the sample selection and distribution of the galaxies
within the Virgo cluster are described in \citet{vSH04}.

\subsection{VATT 1.8m Observations}
UBVRI images of 16 dwarf elliptical galaxies in the Virgo Cluster 
were obtained with the VATT 1.8m on
2003 March  29-31.  The telescope was equipped with CCD26, a thinned Loral 3 2048 $\times$ 2048
CCD with 15\micron~pixels.  CCD26 has a read noise of 5.7 e$^-$ and
a gain of 1.9 e$^-$ per ADU was used.   On-chip binning of 2 $\times$ 2 resulted in 
0.375\arcsec~per pixel and a field of view of 6.4 $\times$ 6.4 arcmin.
The galaxies were centered in the field, but dithers of approximately 1 arcmin
were taken between each exposure to assist with the flat fielding process.

Table \ref{tab:obs} lists the basic parameters of the dE sample and summarizes
the observations.   Unless otherwise noted in Table \ref{tab:obs}, the  VATT 1.8m 
observations consisted of 2 exposures of 540 sec, 420 sec, 300 sec, 240 sec, 
and 300 sec with the U, B, V, R, and I filters, respectively.  
 Complete UBVRI images were obtained for 9 galaxies in the
sample; BVRI images were obtained for an additional 4 galaxies and 3 galaxies
were imaged in only B and R.  The observations were structured so that the B and
R images were obtained on the same night under similar observing conditions; similarly,
the V and I images were taken sequentially.  Thus, the (B-R) and (V-I) colors
for these observations are quite secure, while (B-V) and (U-B) colors have additional 
error propagation terms since the colors are based on the calibrated magnitudes rather
than calibration of the instrumental colors.  The final combined images were
sufficiently deep to permit accurate photometry of the outer regions of the
galaxies.  Typical surface brightness limits for the combined images are
25.8, 26.8, 26.0, 25.6, and 25.0 mag arcsec$^{-2}$, in U, B, V, R, and I, respectively.

All 3 nights were photometric.  Calibration coefficients were derived from observations
of standard stars from \citet{L92} interspersed with the galaxy observations.  Based
on the derived photometric coefficients, the BVRI images have 2\% accuracy while
the U-band images are accurate to 3\%.

\subsection{WIYN 3.5m Observations}

The same sample of galaxies was imaged with 
the WIYN\footnote{The WIYN Observatory is a joint facility of the University 
of Wisconsin-Madison, Indiana University, Yale University, and the National 
Optical Astronomy Observatories.} 
3.5m telescope on 2002 May 12-16.  Poor weather hampered the
usefulness of these observations, but the WIYN images provide confirmation 
for the structural parameters derived from the VATT 1.8m data.  
In particular, the WIYN images have slightly better seeing characteristics 
than the VATT 1.8m images.

The WIYN 3.5m telescope was equipped with the Mini-Mosaic
camera, which consists of two SITe 4096 $\times$ 2048 CCDs with 15\micron~pixels. 
The two CCDs are read-out by 4 amplifiers (2 per CCD), with typical read noises of 
5.5 e$^-$ and gains of 1.4 e$^-$ per ADU.  The pixel scale of Mini-Mo is 
0.14\arcsec~per pixel, which results in a field of view of 9.6 $\times$ 9.6 arcmin 
with a 7.1\arcsec~gap between the CCDs.  In all cases, the telescope pointing was 
offset so that the dwarf elliptical galaxy fit on a single CCD.

The WIYN 3.5m observations consisted of 2 exposures of 600 sec, 600 sec, 300 sec, and 600 sec 
with the B, V, R, and I filters, respectively.
R-band images were obtained for 13 of the 16 dEs in the sample;
additional BVI images were obtained for 6 of the 13 galaxies.
All 5 nights were non-photometric so the limiting surface brightness
varied significantly from exposure to exposure.  Typical image quality for the R-band 
WIYN images is 1-2\arcsec.

\subsection{Data Reduction of Optical Images}
The optical images were reduced and analyzed with the IRAF\footnote{IRAF 
is distributed by the National Optical Astronomy Observatories.} package.
The image reduction included bias subtraction and flat fielding.  
The VATT images were flat fielded to high accuracy using twilight flats.  
The WIYN 3.5m images were first flat fielded using dome flats taken during the
afternoon; since these images had a residual flat field problem, 
a secondary flat field was generated from a combination of
the object frames using standard tasks in the MSCRED package.  Both
the VATT and WIYN I-band images were fringe corrected based on a 
fringe frame generated by taking the median of all I-band galaxy observations
obtained during the night. 

For the WIYN images, data from the two CCDs must
be projected onto a common coordinate system prior to further
data processing.  For each image, the WCS was updated using star 
lists from the USNO catalog and the IRAF tasks MSCZERO and MSCCMATCH.  
The multi-ccd images were then projected into single images using 
the MSCIMAGE task using a common coordinate system for all observations of the
same galaxy.   

For both the VATT and WIYN images, sky values were measured as the 
mode of galaxy-free regions of the images prior to combination of
multiple exposures.  The images were then scaled, aligned, and
averaged together to make a final combined image.
For the non-photometric observations (WIYN), the individual exposures were 
scaled to the image with the most counts; scale factors were determined 
based on the relative intensity of several stars in each frame.  
The VATT data were photometric and required no additional scaling before combination. 

The VATT and WIYN images were processed independently in order
to provide internal consistency checks for the derived
structural parameters.
Since the VATT images are photometric and higher quality
than the WIYN images, the subsequent analysis will focus
on results from the VATT observations.  

\subsection{Optical Spectroscopy}
The optical spectroscopic observations are described in detail in \citet{vSH04}.  
Briefly, optical spectra covering the wavelength ranges of 4800 -- 5700 \AA~ and
8250 -- 8900 \AA~were obtained with the Double Spectrograph on the 5m 
Palomar\footnote{Observations at the Palomar Observatory were made as part of a 
continuing cooperative agreement between Cornell University and the California 
Institute of Technology.}
telescope in 2001 and 2002.  Here, we focus on the ``blue'' spectra, which
were centered on the Mg Ib triplet and include a subset of the spectral regions
appropriate for spectral synthesis using Lick/IDS indices.  The 
observations were obtained on non-photometric, transparent nights.  
Flux calibration was obtained from observations of several spectrophotometric 
standards \citep{oke90} and observations of standard stars from the Lick/IDS
system.   

The long slit spectra were processed following standard practice [see \citet{vSH04}
for complete data reduction details].  After bias subtraction, scattered light 
correction, flat fielding, wavelength rectification, and night sky line
subtraction, the 2-dimensional (2-d) images were flux calibrated.  One-dimensional
(1-d) spectra were then extracted from the 2-d spectra to create spectra
of the integrated light from the galaxy.

The 1-d spectra were Doppler corrected for the systemic velocity of the galaxies.
Since the observed spectra have higher spectral resolution than the stellar spectra that
define the Lick/IDS system, the 1-d spectra were then degraded from 1.6 \AA~resolution
to 8.4 \AA~resolution.  While the Lick indices are based 
on stellar spectra with variable resolution, the spectral region covered by 
these dE galaxy spectra correspond to a region of approximately constant spectral resolution
in the Lick/IDS system \citep{WO97}.   Lick indices were  then measured 
for the galaxies and the Lick standard stars.  The observed spectra
include the spectral regions appropriate for the following
spectral indices: H$\beta$, Fe5015, Mgb, Fe5270, Fe5335, and Fe5406.
In addition, similar to \citet{GGvM03}, we define the average Fe index,
$<Fe>$, as the average of Fe5270 and Fe5335 indices.

The standard star observations were used to confirm that the
indices measured here are consistent with previous measurements.  However, 
since there were only a handful of comparison stars, and the stars observed
comprise a restricted range of spectral types, it was not possible to 
determine a correction to the observed indices to place the measurements 
precisely on the Lick system.  Nonetheless, 
for the few galaxies in common with the \citet{GGvM03} sample, the newly derived
Lick indices are consistent with the previous measurements.  Thus, while 
the derived indices may have some systematic errors, the amplitude of
these errors should be minor.  Furthermore, the {\it relative} measurements
for the dE galaxies in this sample should be secure even if the absolute 
values may be less precise.  

\section{Structural Parameters}

\subsection{Surface photometry}

The surface photometry for each galaxy was derived from ELLIPSE fits
to the final combined images.  Prior to ELLIPSE fitting, the foreground stars
were masked and replaced by the interpolated background of neighboring pixels. 
The surface brightness profiles, position angle, and ellipticity 
are shown as a function of semi-major axis in Figure \ref{fig:surf}.
Since the morphological parameters were similar for all filters,
the apertures defined by the R-band fit were used for subsequent analysis
of the surface brightness profiles and surface colors.

The morphological parameters  are summarized in Table \ref{tab:results}.
Listed in Table \ref{tab:results} is the semi-major axis at which the B-band
surface brightness drops to 25 mag arcsec$^{-2}$, R$_{25}$, and the
average ellipticity and position angle from the isophotal fits of
the R-band images for ellipses with semi-major axes between 5\arcsec~and R$_{25}$.
Apertures larger than 5\arcsec~were used for these average values since
several of the galaxies in this sample have bright nuclear regions.
Also listed in Table \ref{tab:results} are the B-band apparent magnitudes
at which the curves of growth level off and the observed optical
colors as measured within the R$_{25}$ aperture.  As illustrated in 
Figure \ref{fig:surf}, the surface colors and aperture colors reach
similar values near R$_{25}$ so these values were chosen as the 
most representative galaxy color.  Also tabulated in Table \ref{tab:results}
are estimates of the foreground Galactic extinction as calculated 
from the dust maps of \citet{SFD98}.

The images were examined for residual disky or boxy isophotes by
subtracting the model surface photometry from the image.  Eleven of
the 16 galaxies had residual structure that was readily apparent
in the residual images; the presence of disky or boxy isophotes is 
tabulated in Table \ref{tab:results}.
Deviations from the best-fitting ellipse as measured by B$_4$ are shown
in Figure \ref{fig:surf}. 

As previously noted by \citet{GGvM03}, there appears to be
little correlation between the presence of disky or boxy residuals 
and the kinematic properties of the galaxies.  As an indication
of the kinematic properties of these galaxies, the rotation curve 
slope from \citet{vSH04} is listed in Table \ref{tab:results}.  
Due to the limited spatial extent of the stellar kinematic observations, 
we list the rotation curve slope rather than a maximum velocity width;
galaxies with rotation curve slopes greater than 20 km s$^{-1}$ 
kpc$^{-1}$ have significant rotation.  Based on the kinematic
observations,  VCC 178, 437, 543, 965, 990, 1036, 1743, 1857, and 2019
form the subset of rotation dominated dEs in this sample.

Table \ref{tab:results2} summarizes the derived photometric properties of 
the dwarf elliptical galaxies in this sample.  
The B-band absolute magnitude was calculated from
the apparent magnitude assuming all galaxies are at the Virgo Cluster distance
of 16.1 Mpc \citep{Ke00} and correcting for Galactic extinction.  
Also listed are the effective radius
and effective surface brightness in B-band.  For all of the following
except the exponential parameters, size parameters are quoted
in geometric mean radii where r = $\sqrt{ab}$; for the exponential parameters,
the semi-major axis is adopted as the radius.
The effective radius tabulated in Table \ref{tab:results2} is the half-light radius
based on the apparent magnitude listed in Table \ref{tab:results};
the effective surface brightness is then defined as the average surface brightness
within the R$_{\rm eff}$ elliptical aperture.  We note that
while these values are model independent, 
they can be sensitive to systematic errors introduced
by shallow imaging observations
since the half-light radii are defined based on the observed
apparent magnitudes.

Several of the galaxies in this sample have surface photometry reported
in the literature.  Recent observations with partially overlapping samples
include VCC 543, 917, 1036,
1261, and 1308 reported in \citet{GGvM03}; VCC 965, 1036, 1122, and 1308
reported in \citet{P02}; VCC 1036 and 1261  reported in \citet{BBJ03}; and VCC 543
and 1743 reported in \citet{D97}.  In general, there is reasonable agreement
between the literature values and those reported here, particularly
given the variety of filters and telescopes used by these studies.
However, there are notable differences between the values reported
here and the magnitudes and effective radii listed in \citet{GGvM03}.
With the exception of VCC 1261,
it is likely that many of these differences can be attributed to 
differing image depths and to incomplete surface photometry
as a result of the restricted field-of-view of HST imaging observations.  
For VCC 1261, the large discrepancy between the effective radius reported
here and that reported in \citet{GGvM03} (more than a factor of 2) cannot
be reconciled by mere observational differences; the present value
agrees with the surface photometry of VCC 1261 reported in \citet{BBJ03}.

\subsubsection{Exponential Fits}

To enable comparison with galaxies of other morphological types,
the outer isophotes (r $>$ 10\arcsec) of the B-band surface 
brightness profiles were fit to an exponential model:
\begin{equation}
\mu(r) = \mu^0 + 1.086 r / \alpha
\end{equation}
where $\mu(r)$ is the observed surface brightness at semi-major axis $r$,
$\mu^0$ is the extrapolated central surface brightness, and $\alpha$
is the exponential scale length. 
The face--on central surface brightness, $\mu_B^{0,c}$,
was calculated by applying a Galactic extinction correction and a
line--of--sight correction of $-2.5~{\rm log~(cos}~i)$, where $i$ is the inclination
derived from the observed axial ratios and a thin disk approximation. 
While an exponential distribution does not necessarily
imply a disk-like stellar distribution, the simplicity of this
functional form permits a robust analysis of the light distribution
of the outer regions of a galaxy, regardless of the actual shape of
the galaxy (disk or spheroid).
The exponential fits enable a direct comparison between the 
structural parameters of dE and dI galaxies (Figure \ref{fig:struct}).  
The top panel of Figure \ref{fig:struct}
shows the scale length as a function of absolute magnitude while the
bottom panel shows the face--on central surface brightness as a function
of absolute magnitude.  For comparison, the structural parameters for
dwarf irregular galaxies from \citet{vZ00} are also shown.  Despite the
fact that this comparison sample may not be ideal, since the dIs 
were selected to be isolated galaxies and the dEs are cluster members, these
figures confirm that the structural parameters of dIs and dEs are 
quite similar \citep[as first discussed in][]{LF83}.   
In particular, both the dI and dE samples contain low surface brightness 
galaxies with typical central surface brightnesses between 22 to 24 mag 
arcsec$^{-2}$ and scale lengths of only a few kpc.

\subsubsection{S\'ersic Fits}

While the parameters for an exponential fit are sufficient to represent
the stellar distribution in the outer galaxy, 
a more generalized surface brightness profile, such
as the S\'ersic function \citep{S68}, can provide additional insight into the stellar
distribution.  The surface brightness profiles were fit to a S\'ersic function of
the form: I = I$_0$ e$^{-(r/r_0)^n}$.  In magnitude space, this corresponds to
\begin{equation}
\mu = \mu_s + 1.0857(r/r_s)^n
\end{equation}
where $\mu_s$ is the central surface brightness, r$_{\rm s}$ is a
size parameter, and $n$ is a shape parameter.  In this form, n=0.25
corresponds to an R$^{1/4}$ law distribution, while n=1 corresponds
to an exponential (disk) distribution.   The surface brightness
profiles were fit for geometric mean radii between 1\arcsec~and the
radius at which the R-band surface brightness reached 25 mag arcsec$^{-2}$.
Since the derived parameters were similar regardless of filter choice,
only the R-band fits are listed in Table \ref{tab:results2} and
shown in Figure \ref{fig:surf}.  In general, the
surface brightness profiles were well fit by the S\'ersic function,
with shape parameters, n, between 0.5 and 1.  
However, the S\'ersic parameterization proved to be quite sensitive to choice
of fitting radii.  Thus, while errors are listed in Table \ref{tab:results2},
equally valid solutions can be found with parameterizations that are
outside of the range apparently permitted by these formal errors.

Several of the galaxies in this sample have S\'ersic parameters listed in the literature.
The present results agree with the literature values for
VCC 1036 and 1261 \citep{BBJ03}, 
VCC 1308 \citep{GGvM03},
and VCC 1743 \citep{D97}.
However, when overlayed with the observed surface brightness profile,
the S\'ersic parameters for VCC 543, 917, and 1035
from \citet{GGvM03} yielded good fits only to the inner regions
of these galaxies (as mentioned above, their values for VCC 1261 do not 
match the surface brightness profiles obtained here).  Similarly, the 
parameters listed in \citet{D97} for VCC 543 are only applicable to the inner profile. 
The disagreement between the present study and the literature 
is not surprising since S\'ersic fits are sensitive to the choice of fitting radii,
and the previous analysis often restricted the fits to the inner 20-30\arcsec.  

The results of S\'ersic fits for dwarf elliptical galaxies in the Virgo Cluster
are shown in Figure \ref{fig:sersic}.  Figure \ref{fig:sersic} includes all dE galaxies 
fainter than M$_B$ of -17.5 with published kinematic data \citep[e.g., the sample includes
galaxies compiled from][]{PGCSBG02,GGvM03}.
Structural parameters for many of the galaxies in \citet{PGCSBG02} and \citet{GGvM03}
were compiled from \citet{BBJ03}.
For those galaxies in \citet{GGvM03}
without published B-band effective surface brightnesses, the published 
V-band effective surface brightness was converted to an approximate B-band value by
assuming that the typical (B-V) color of a Virgo dE is 0.77 (see Section 4.2).
While Figure \ref{fig:sersic} indicates that
there is a general trend of larger shape parameters (n) for fainter
dwarf ellipticals, and for brighter central surface brightnesses for brighter
galaxies, the rotating and non-rotating dwarf elliptical galaxies cannot be
distinguished in any of these figures.  In fact, the kinematic samples
are well mixed in terms of shape, surface brightness, and size.

In summary, there appear to be no strong morphological differences between
rotating and non-rotating dwarf elliptical galaxies in the Virgo cluster.
Underlying disky and boxy isophotes are seen in both kinematic samples
\citep[see also][]{GGvM03}.  The structural parameters are similar as well.
In the next section, we investigate whether there are stellar population
differences between the rotating and non-rotating dwarf elliptical galaxies.

\section{Optical Colors and Color Gradients of Dwarf Elliptical Galaxies}

\subsection{Optical Color Gradients}
As illustrated in Figure \ref{fig:surf}, many of the dwarf elliptical galaxies
in this sample have a slight color gradient.  However, this trend
does not appear 
to be correlated with the presence of a nucleated region; several of
the nucleated dEs have no color gradients while several of the
non-nucleated dEs do have a color gradient.  Color gradients are commonly
seen in dEs, usually in the sense that the outer regions are redder 
than the inner regions \citep{VVLS88,JBF00,BBJ03}.  In this sample,
VCC 965, 1743, 1857, and 2019 have color gradients in this expected direction.  
However, approximately one half of the current sample have color gradients in 
the opposite sense; the inner regions are redder than the outer regions 
of VCC 178, 543, 917, 990, 1036, 1122, 2050. 

A color gradient implies differing star formation histories, or differing
metal content, in the inner and outer regions of the galaxy.  
It is likely that both of these effects
are relevant to the interpretation of observed colors and 
stellar populations of Virgo dEs.   
For example,
a galaxy may develop a color gradient if the outer gas is preferentially
stripped off as it falls into the Virgo cluster.  In this scenario,
the inner region of the galaxy may continue to have active star formation
while the outer regions are reduced to an aging stellar population (redder colors).
Such a scenario easily explains the observed color gradients in most dwarf
elliptical galaxies (redder in the outskirts).  Alternatively, a color gradient
in the opposite sense can be created if the inner region retains a larger
fraction of its enriched material; if there is a metallicity gradient, 
the more metal-rich stars (inner regions) will be redder than the 
metal-poor stars (outskirts).  The relative importance of both of these
effects will depend on the galaxy's detailed star formation history and the
interface of the galaxy with the intracluster medium.

As shown in Figure \ref{fig:rot}, the presence of a color gradient
does not correlate with the kinematic properties of the dEs.  Galaxies
with a blue core (VCC 965 and 2019) or red core (VCC 543, 917, 990, 1036, and 2050)
are equally likely to be rotation dominated as galaxies with no strong color gradient.
Also shown in Figure \ref{fig:rot} is the luminosity-line width correlation
for spiral galaxies \citep{TP00} and for field dwarf irregular galaxies 
compiled in \citet{vZ01}.  As discussed in \citet{vSH04}, the observed dE
rotation velocities are lower limits due to observational constraints;
reasonable extrapolation of the observed rotation curves result in
linewidths comparable to galaxies that are 1-2 magnitudes brighter than
the current dE population.  This fading estimate is in reasonable agreement
with that expected if dEs are evolved dIs \citep[e.g.,][]{BMCM86}.  Nonetheless,
it is quite remarkable that the observed (un-extrapolated) rotation velocities
show excellent agreement with the bright galaxy Tully-Fisher relation.

\subsection{Optical Colors}

The presence of a color gradient complicates the definition of a global galaxy
color.  To avoid systematic trends due to nuclear contamination of the colors,
the colors tabulated in Table \ref{tab:results} are measured within the 
B-band 25 mag arcsec$^{-2}$ isophotal aperture; this aperture color was 
selected to be representative of the true color of the galaxies since the aperture and
surface colors are quite similar at this radius (see Figure \ref{fig:surf}).  
Table \ref{tab:results2} lists the observed optical colors corrected for Galactic
extinction.  The optical colors are remarkably similar for all of the
dEs in this sample.  The median colors are 0.24 $\pm$ 0.03 in (U--B), 
0.77 $\pm$ 0.02 in (B--V), 
1.26 $\pm$ 0.05 in (B--R),
0.48 $\pm$ 0.01 in (V--R),
and 1.02 $\pm$ 0.03 in (V--I).
Observations of low luminosity elliptical galaxies yield similar
colors for galaxies in the same absolute magnitude range \citep{PBKA93}.

The optical colors are shown in Figure \ref{fig:evol};  
as expected, the dwarf elliptical galaxies are red systems in
all color combinations.
For comparison, the evolutionary tracks for 3 possible star formation 
histories are shown in Figure \ref{fig:evol}.  The evolutionary
tracks are obtained from the simple stellar population models
of \citet{BC03} assuming a mean metallicity of 1/5 solar; 
the low metallicity stellar population models were selected
for this figure because low luminosity galaxies are usually low metallicity 
systems \citep[e.g.,][]{SKH89,RM95}.  The \citet{BC03} galaxy evolution code
is an improved version of the composite stellar population models
originally described in \citet{BC93}.  
As in \citet{BC93}, the mono--metallicity composite stellar populations 
are built through the synthesis of stellar evolutionary tracks \citep[e.g.,][]{LCB97}.   
The version of the galaxy evolution code run here includes options for sub-- and 
super--solar metallicity stellar populations,
in addition to allowing the user to vary the star formation history and the initial mass
function. For simplicity, a \citet{S55} IMF was adopted for all of the models considered here;
a top-- or bottom--heavy IMF will result in minor variations in the evolutionary tracks, but
relative conclusions  should be robust with regard to the choice of IMF.

The three fiducial star formation histories shown in Figure \ref{fig:evol} 
include (1) a simple burst of star formation that lasts
10 Myr, (2) an extended burst of star formation with an e-folding time
of 1 Gyr ($\tau = 1$), and (3) a constant star formation rate model.
As expected, the constant star formation rate model can be ruled out for
the Virgo dEs since these galaxies have no current star formation activity; 
this model is appropriate for dwarf irregular galaxies which have
quasi-continuous star formation activity \citep{vZ01}.  The observed
colors are consistent with both fiducial star burst models (short and
extended starbursts).  In both cases, the observed colors are indicative
of an older stellar population with an age of approximately 6 Gyr.
While there is some ambiguity in this age estimate
due both to the well known age-metallicity degeneracy and to possible
different star formation histories, it is clear that the observed colors
are indicative of a stellar population which has had little-to-no star
formation activity within the last few Gyr.
Thus, while the observed broad band colors do not permit a detailed
analysis of the past star formation activity in these galaxies,
they do indicate that none of the galaxies in this sample are
{\it recently} stripped dwarf irregular galaxies.  

Further, the small range of observed colors for the galaxies in this
sample indicates that the dominant stellar population is
quite similar for both the rotating and non-rotating
dwarf elliptical galaxies.  As illustrated in Figure \ref{fig:colmag},
much of the scatter in the observed colors can be attributed
to the well known color-magnitude relation for dwarf elliptical
galaxies \citep[e.g.,][]{C83}.  In general, the more luminous galaxies 
in this sample are slightly redder, as would be expected if the more
luminous galaxies in the sample are also more metal-rich \citep[e.g.,][]{SKH89}. 
The two significant outliers to this
relation are VCC 1857 and VCC 1261, both of which are bluer than
one might expect for their absolute magnitude.  Interestingly,
neither of these galaxies have a significant color gradient,
so their anomalous colors cannot be explained by nuclear
contamination.  We note also that the present results for
VCC 1261 are consistent with the B-R color reported in
\citet{BBJ03} for this galaxy.   It is suggestive that VCC 1857
is not only bluer than expected, but also has slightly different
structural parameters than the other Virgo dEs; this galaxy
is the best candidate in this sample for a recently stripped
dI.  Nonetheless, its optical colors are still consistent with
an older stellar population, indicating that the last
star formation activity occurred several Gyr ago.  Unfortunately, the kinematic
observations of VCC 1857 were inconclusive; while the derived rotation
curve indicates that VCC 1857 may have a significant rotational component,
the rotation curve was poorly constrained because of the galaxy's
low surface brightness nature.  In contrast, the other anomalous galaxy,
 VCC 1261, is the one galaxy in the sample with clear absence of a rotational 
component along its major axis.  

Using these two outliers as indicative examples, the relative colors of 
the Virgo dEs do not appear to be related to their kinematic signatures.
Rather, the small range of observed colors for the galaxies in this sample
indicates that the ages of the dominant stellar population are approximately
the same for both rotating and non-rotating dwarf elliptical galaxies.
We note, however, that broad band colors do not permit a detailed analysis
of the past star formation activity in these galaxies since the observed
colors are consistent with both a short burst and with a more extended
star formation history.  We explore additional
constraints on the past star formation activity by examining
the chemical signatures of the dominant stellar population 
in the next section.

\section{Age and Metallicity: Spectroscopic Evidence}

The results of the spectral analysis are shown in Figure \ref{fig:alpha}.
The expected Lick indices for three possible [$\alpha$/Fe] ratios
are shown based on the models of \citet{TMB03}.
Unlike giant elliptical galaxies, the majority
of dwarf elliptical galaxies in Virgo appear to have solar or sub-solar 
[$\alpha$/Fe] ratios.  Similar results are found by \citet{GGvM03} 
and \citet{TBHMG03} for their samples of Virgo dwarf elliptical galaxies;   
in addition, similar [$\alpha$/Fe] ratios are seen in stellar abundances of nearby
dwarf spheroidals \citep[e.g.,][]{SCS01,SVTPHK03}.
Since Type II supernova produce the majority of $\alpha$--rich elements, the observed 
[$\alpha$/Fe] ratios trace the timescale of star formation
activity in each galaxy \citep{GW91}.  In particular, super-solar [$\alpha$/Fe] 
indicates rapid enrichment from Type II supernova and thus implies that the galaxy
has undergone a short burst of star formation activity; most
giant elliptical galaxies have super-solar [$\alpha$/Fe]
\citep[e.g.,][]{TFWG00}.  In contrast, the solar and sub-solar
abundance ratios for dEs indicate slow chemical enrichment, or a more quiescent
star formation history.  Thus, the dominant stellar populations in Virgo
dEs may not be post-burst populations.  Rather, previous star formation
activity appears to have occurred on more extended timescales in the
low mass galaxies; these and similar results \citep[e.g.,][]{GW94,M98}
rule out various theoretical models for dE formation, including the possibility
that dEs and dSphs form all 
stars prior to reionization \citep[e.g.,][]{BL99}.   
The observed [$\alpha$/Fe] ratios are consistent with the idea that 
Virgo dEs are formed through the stripping of dIs; if star formation activity 
ceases when the ISM is removed from the low mass system, the galaxy will evolve
into a quiescent, red galaxy with structural and kinematic parameters similar
to the progenitor dI, and with a chemical enrichment history representative
of a more continuous star formation history.  

Comparison of the H$\beta$ and Mgb indices provides an indication
of the age of the dominant stellar population in the dEs (Figure \ref{fig:metals}).  
Also shown in Figure \ref{fig:metals} are the stellar
population models of \citet{TMB03} for solar [$\alpha$/Fe].
As expected from the optical colors (Section 4), the stellar
populations are both metal--poor and evolved.  Based on
these observations and stellar population models, Virgo dEs
have typical ages of 5-7 Gyr and metal abundances between 1/20
and 1/3 of solar.
Similar results for Virgo dEs were 
also obtained by \citet{GGvM03}.  The derived age and metallicity
estimates for these galaxies are also in remarkably good agreement
with those obtained from the broad band optical images.
As found with the optical colors, there is no clear separation
of the kinematic samples in either age or metallicity of the
dominant stellar population.
Thus, both stellar population models and spectral synthesis indicate
that Virgo dEs contain evolved low metallicity stellar populations with typical
ages of a several Gyr and solar or sub-solar [$\alpha$/Fe].

\section{Discussion}

The derived structural parameters and metal enrichment histories 
indicate that the Virgo dEs are similar in nature to field dwarf 
irregular galaxies with the exception of the age of the dominant
stellar population.   Thus, the current
observations are consistent with the idea that some cluster dEs are
the remnants of dIs which have been stripped of their ISM.
Taken at face value, the derived mean stellar population ages indicate
that the last major star formation episode in the Virgo dEs occurred
5-7 Gyr ago, or at $z > 0.5$.   However, due to the inherent ambiguities
in stellar population models, it is difficult to determine exactly how long
it has been since star formation occurred in these galaxies.   For 
example, the observed colors are not only consistent with the two burst 
models discussed above (extended starburst and short burst), but also with a galaxy
which has had a constant star formation rate which is abruptly
truncated 3 Gyr ago ($z \sim 0.3$).  Unfortunately, the Virgo dwarfs are
too distant to produce detailed star formation histories as is possible
within the Local Group \citep[e.g.,][]{STCDSGDM03}.  Nonetheless, all of these models
indicate that the Virgo dEs have not had substantial star formation activity
within the last few Gyr.  

Furthermore, the observed [$\alpha$/Fe] ratios are solar or sub-solar,
indicating that the stellar mass of the cluster dEs has been built
gradually, rather than through a major burst of star formation.  
Thus, not only do low mass cluster galaxies have different structural 
properties than their giant cousins, their stellar populations and star 
formation histories appear to be substantially different.  While 
both giant and dwarf elliptical galaxies appear to have strong 
environmental dependences on their formation and evolution,
it is likely that low mass cluster galaxies follow different evolutionary 
paths than their giant counterparts.  For example, interactions with the
cluster potential and the ICM may be the key to formation of
dEs, while galaxy-galaxy interactions may play a major role in the 
evolution of giant ellipticals.

\subsection{Are the Majority of Virgo dEs Produced by the Infall and Stripping of dIs?}

Since the early recognition of similar structural parameters between
dE and dI galaxies \citep[e.g.,][]{LF83,K85}, it has often been suggested
that dEs could simply be dIs stripped of their gas and left to fade.
The extensive study of the Virgo cluster resulting in the VCC
\citep[c.f.,][]{VCC,SBT85,BTS87} confirmed the strong overlap
between these two populations, but led \citet{B85} to conclude that the 
dEs are not formed from dIs due to the apparent lack of a bright progenitor
dI population and to the presence of bright nuclei in a large fraction of dEs. 
While \citet{B85} was careful to note that his arguments applied only to the
brightest dEs, he argued that if one could not explain the entire
dE population with stripping then it did not make sense to invoke this
process since it is inelegant to explain the well defined sequence
in L, $\alpha$, $\mu$ space with more than one process.

In \citet{vSH04}, we proposed that Virgo dEs are formed
through multiple processes.  First, it is likely that the 
most massive galaxies near the core of the cluster formed 
with a coterie of satellite galaxies, which soon became dE galaxies.  
However, we also know that the Virgo cluster is still forming
today, and that field galaxies and groups of galaxies are falling into
the cluster \citep{TS84,BTS87}.  If the groups of galaxies that 
later fell into the Virgo cluster were similar to our Local Group, 
then they likely consisted of roughly equal numbers of dE and 
dI galaxies before entering the cluster environment.  
In addition, the infalling dI galaxies are likely to be affected
by the cluster environment and may lose their ISM through 
ram pressure stripping or through interactions with other
galaxies.  It therefore seems unavoidable that there are
at least three different origins for the Virgo dEs ({\sl in situ} formation
near massive elliptical galaxies, infalling dEs that were once part of Local Group
analogs, and converted infalling dIs).  
Thus, the narrow locus formed by dEs in structural parameter space
may better be interpreted as the result of an inevitable structural 
form as opposed to a uniform process of creation.  In other words, 
the uniformity of dE/dI properties may be due more to the uniformity of 
their dark matter halos \citep{NFW97}, rather than a single formation mechanism.

   Additional arguments have been made against the stripping of infalling
dIs to form the Virgo dE population.  
\citet{GH89} reason that the scarcity of dwarf galaxies with blue
colors and low HI content argues against extensive stripping of dI
systems in the last $\sim$Gyr.  
However, recent observations have identified several likely candidates
as recently stripped dIs.  The best case
for stripping of a dwarf galaxy in process is that of the HI cloud
between NGC 4472 and UGC 7636 discovered by \citet{STE87}.  Subsequent
observations of this system have shown that the morphologies of the
optical galaxy and the HI cloud, although separated by 15 kpc, are 
``remarkably similar'' \citep{MSHJ94} and that the ISM
abundance of the cloud agrees with that expected for the
luminosity of UGC 7636 \citep{LRM00} - essentially clinching the
proposal that this is a recent stripping event.  \citet{LKG97} found
blue star clusters with ages up to 10$^8$ yrs associated with both
the optical galaxy and the HI cloud and this provides a minimum time
since the stripping event.  In addition to UGC~7636, IC~3475 has been 
identified as a recently stripped irregular galaxy \citep{VTVL86}, 
IC~3365 appears to be a dI caught in the act of being stripped \citep{SBMW87},
and VCC~882 is a nucleated dE that from a long trail of dust appears to
have recently been stripped of its ISM \citep{EECF00}.  

   Anecdotal evidence allows us to be sure that the process of ram pressure 
stripping is viable, but a statistical approach is required to address the
point made by \citet{GH89}.  From a large
sample of HI observations of Virgo dIs,  \citet{HHSS85} concluded that 
the HI depletion of dIs is only moderate.  However, this was based on
comparing M$_{\rm H}$/L ratios of dIs to those of spirals.  The means of 0.2
for the bright Virgo dIs and 0.5 for the faint Virgo dIs are both significantly smaller
than the typical M$_{\rm H}$/L of 1 for field dIs \citep[e.g.,][]{FT75,vHG95,S96}.
From this, one can conclude that a significant fraction of Virgo dIs
are showing the effects of the cluster environment.  \citet{LMR03} have
revisited this question, and a comparison of Virgo dIs with field dIs
show the majority of Virgo dIs in their sample have HI deficiencies.  Finally, from our
color and spectroscopic dating of Virgo dEs, it is clear that the majority
have been dEs for several Gyr.  If most of the conversions happened 
long ago, a low transformation rate today might only imply that the 
process is slowing down, rather than excluding it altogether.
In fact, the argument is based on the assumption of
a constant rate of creation of dEs over the Hubble time.  Given the
stochastic nature of infall, a variable rate of infall would also be
consistent with a low rate of dE creation in the present epoch and a
large population of dEs formed from stripped dIs.

Arguments against conversion from dIs to dEs by ram pressure stripping
have also been made on theoretical grounds.
\citet{FB94} revisit this question and offer a rough calculation of the
time scale for ram pressure stripping and note that it is unreasonably
long for distances greater than 300 kpc from M87 (which excludes only the very 
core of the Virgo cluster).  They conclude that ``stripping was 
probably not the dominant gas-removal mechanism'' in the Virgo cluster.
However, they note that a more realistic treatment of the ISM in a
dI might yield significantly shorter gas removal times.
The empirical counter example to this theoretical calculation comes in
the form of HI observations of the spiral galaxy NGC~4522
\citep{KvGV04}.  They show this 0.5 L$^{\star}$ galaxy to be undergoing 
significant stripping at a projected distance of $\sim$ 800 kpc from M~87.
The stripping of the HI is complete to significantly higher mass
surface densities than seen in most dwarf galaxies, and thus any
gas rich dwarf in this environment is likely to be stripped.

Binggeli's \markcite{B85} (1985) argument concerning the lack of a progenitor population
for nucleated dEs remains as a possible constraint on the number of
Virgo dEs which are created as a result of stripping of an infalling
dI.  Unfortunately, without a consensus view on the formation of the
nuclei of nucleated dEs \citep[see, e.g.,][]{OL00, BBJ00}, it is
difficult to judge the strength of this constraint.
The fact that the nucleated dEs are more strongly clustered
than the non-nucleated dEs \citep[e.g.,][]{BTS87} 
may be an indication that they are part of an
{\it in situ} population of galaxies formed as satellites of the original
cluster galaxies.  Based on specific globular cluster frequencies,
\citet{MLFSW98} propose that the non-nucleated dEs are the more
likely dEs to form as a result of stripping.  However, in this regard,
note that 3 of the 5 galaxies in the \citet{vSH04} sample with
strong rotational support (v/$\sigma$ $>$ 1) are nucleated dEs.
Alternatively, it is possible that nucleated dEs
could be formed from dIs with large central starbursts \citep{DP88}.
Further study of their stellar populations and structure are needed
to understand fully the origin of the nuclei and thus to 
determine if stripped dIs can evolve into nucleated dwarf ellipticals. 

Finally, a possible concern for dI to dE evolution through ram pressure
stripping is a possible disparity in baryon-to-dark matter ratios 
in these low mass galaxies.  
While the characteristics of the dark matter halos of the Local Group
dSph galaxies are currently a matter of debate
\citep[cf.,][]{KWEGF02, PPAO02, SWTS02, PMSPOL03}, it appears 
that dSph may have lower dark matter-to-luminosity ratios than similar
luminosity dwarf irregular galaxies \citep[e.g.,][]{M98}.
However, such results should be viewed with some caution since dark matter halos are
measured in different ways for rotating and non-rotating systems \citep{M98}.
Thus, at this time, possible discrepancies in the masses of dark matter halos
and baryon-to-dark matter ratios of non-rotating Virgo dEs 
are not sufficiently robust to exclude ram pressure stripping
as a viable evolutionary pathway from dI to dE.
Further, as shown in Figure \ref{fig:rot}, the rotating 
Virgo dEs have maximum rotation velocities comparable to 
similar luminosity dIs, suggesting that at least this subset
of dE galaxies may have similar baryon-to-dark matter ratios as
field dIs.  

    In summary, after reviewing the historical comparisons of the dE and 
dI populations in Virgo which have been used to rule out the creation of 
the Virgo dEs by infalling dIs, we find no observational evidence which
rules out the creation of a majority of the Virgo dEs by the stripping of
gas from dIs.  We note, however, that if ram pressure stripping is a common
mechanism for transforming dIs to dEs, there remains a fundamental
difficulty of identifying the population of recently stripped dIs.
In all likelihood, the recently stripped dIs are not (yet) classified as dEs 
(UGC 7636 is dI).  In fact, any young stellar population must have already faded
by the time a galaxy is classified as a dE since the primary morphological
criterion is that dEs have smooth and regular isophotes.  Thus, it should
not be surprising that all the dEs in this sample have similar optical
colors.  By the time a galaxy is classified as a dE, its stellar population
has aged sufficiently ($\sim$2 Gyr) that the observed colors are red and
it is no longer possible to determine a precise time since the last
star formation episode.  In analogy with the Local Group dSph, however,
it is likely that the detailed star formation histories of the Virgo
dEs (as revealed by color magnitude diagrams) will differ
substantially even though the integrated colors are similar.

\subsection{Are the Rotating dEs and Non-Rotating dEs Formed Differently?}

The optical images and spectroscopy presented here indicate that
there are no obvious differences between the stellar populations of rotating 
and non-rotating dwarf elliptical galaxies.   In \citet{vSH04}, we 
proposed the possibility that non-rotating dEs may
have entered the cluster environment as dEs, while rotating 
dEs may have been converted from dIs while entering the cluster 
environment.  If this scenario is correct, then one would expect, on 
average, that the non-rotating dEs would have older
stellar populations.  Unfortunately, the age resolution afforded by
the present observations and the sample size are insufficient to provide 
a clear test.  There is the additional complication that a rotationally 
supported dE which is a result of a ram pressure stripping event may
follow a trajectory in which a close encounter significantly alters its
kinematics \citep[e.g.,][]{Me01b}.  Thus, even if the rotating and non-rotating
dEs are entering the cluster environment in similar distributions,
the distributions of the observed rotating and non-rotating 
dEs could be significantly different.   

The best way to sort out this question observationally may be to 
study the population of recently stripped galaxies and those galaxies
being actively stripped today.  An unbiased HI survey of the Virgo
cluster environment may reveal a large number of HI clouds similar
to the one associated with UGC~7636.  If so, then the distribution 
(in both real space and velocity space) of these objects and their
association with rotating or non-rotating galaxies will offer
significant constraints on all scenarios accounting for the dE
population in Virgo.  The new seven feed L-Band array (ALFA) at Arecibo
Observatory is well suited to such a blind neutral hydrogen
survey.  With a relatively small beam (3 arcmin), ALFA observations
of the Virgo cluster will have sufficient spatial resolution and
sensitivity to identify recently stripped HI clouds and gas--poor
dwarf irregular galaxies within this high density region; such observations
will improve significantly our understanding of the population
and location of recently stripped dIs and their connection to the
dwarf elliptical population.  

\subsection{Are Intermediate-z Faint Blue Galaxies the Progenitors of Dwarf Ellipticals?}

The abundant dE population in the local universe implies the
existence of a large population of star forming dwarf galaxies
at intermediate redshifts, corresponding to progenitors
of the local dE population.  The detectability and the
nature of this population depend entirely on the evolutionary
histories of dEs.  If dEs build stellar mass rapidly, they
could appear as luminous objects ($> L^{\star}$) during
their formation and then fade by $> 3$ magnitudes to the present
epoch.  Such ``starburst'' scenarios are favored to explain
the rapidly evolving population of faint blue galaxies at intermediate
redshift \citep[e.g.][]{BR92}.  However, the observed [$\alpha$/Fe]
indicates that dEs build their stellar mass more slowly.
If most dEs form from dIs, with their roughly constant star formation histories 
punctuated by, at most, moderate bursts of star formation, the forming 
dEs may be only 1-2 magnitudes brighter than those at the present
epoch.  Thus, with expected magnitudes M$_{\rm B} \geq -18$ 
(m$_{\rm R} > 24$ at $z=0.5$), the population of dE progenitors
will be rare in most extragalactic surveys.   

We caution, however, that integrated colors and elemental
abundance ratios provide only modest constraints on the detailed
star formation history of a galaxy since a modest star burst, or period of 
quiescence, can be masked by subsequent star formation episodes. 
Thus, while the present observations
are suggestive that the dI to dE transition does not require
a starburst episode, observations of the resolved stellar populations 
will be necessary to determine the detailed star formation history
of dwarf elliptical galaxies.  
Such observations are beyond the capabilities of current optical 
telescopes, but the next generation of ground based telescopes may make this 
a realistic goal even at the distance of the Virgo cluster.  
Such observations are crucial for our understanding and interpretation 
of galaxy counts at moderate to high redshift.

\section{Conclusions}

We present optical images and spectral synthesis of 16 dwarf elliptical galaxies
in the Virgo cluster.  The major results of these observations are
as follows:

(1) There appear to be no major morphological differences between
rotating and non-rotating dwarf elliptical galaxies.  Underlying boxy
and disky isophotes are seen in both kinematic samples.  As first discussed
by \citet{LF83} and \citet{K85}, the cluster dEs have similar structural 
parameters as field dIs.

(2) The global optical colors are red, with median values for
the sample of 
0.24 $\pm$ 0.03 in (U--B),
0.77 $\pm$ 0.02 in (B--V),
1.26 $\pm$ 0.05 in (B--R),
0.48 $\pm$ 0.01 in (V--R),
and 1.02 $\pm$ 0.03 in (V--I).

(3)  Stellar population models and spectral synthesis indicate that
the dominant stellar populations have ages of approximately 5-7 Gyr.
Based on Lick indices and simple stellar population models, the 
derived [$\alpha$/Fe] ratios are sub-solar to solar,
indicating a more gradual chemical enrichment history for dEs
as compared to giant elliptical galaxies in the Virgo Cluster.

(4) We argue that it is likely that several different physical mechanisms
played a significant role in the production of the Virgo cluster dE galaxies
including {\sl in situ} formation, infall of dEs that were once part of
Local Group analogs, and the conversion of infalling dIs.
 The present observations support the hypothesis
that a large fraction of the Virgo cluster dEs are formed by ram pressure
stripping of gas from infalling dIs.

\acknowledgements
This research has made use of the NASA/IPAC Extragalactic Database (NED)
which is operated by the Jet Propulsion Laboratory, California Institute
of Technology, under contract with the National Aeronautics and Space
Administration. We thank the VATT time allocation committee for awarding
TBS time to complete this project.
 LvZ acknowledges partial support by Indiana University.
Support for EJB was provided by NASA
through Hubble Fellowship grant \#HST-HF-01135.01 awarded by the Space
Telescope Science Institute, which is operated by the Association of
Universities for Research in Astronomy, Inc., for NASA, under contract
NAS 5-26555.  EDS is grateful for partial support from a 
NASA LTSARP grant No. NAG5-9221 and the University of Minnesota.

\begin{table}
\dummytable\label{tab:obs}
%this table contains the list of observations
\end{table}

\begin{table}
\dummytable\label{tab:results}
%this table contains the imaging parameters
\end{table}

\begin{table}
\dummytable\label{tab:results2}
%this table contains the surface photometry parameters
\end{table}

\newpage

\psfig{figure=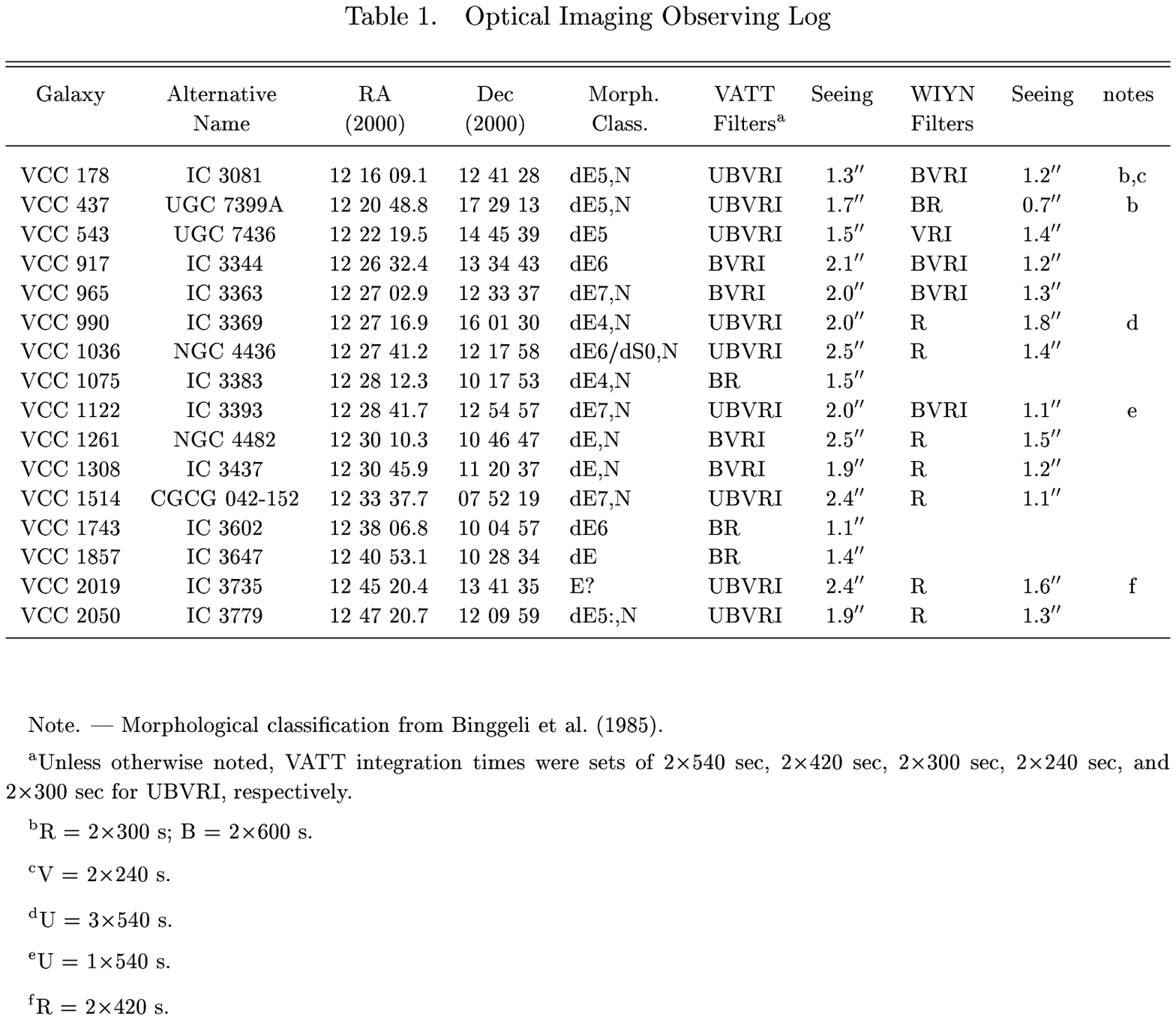,height=25.cm}

\psfig{figure=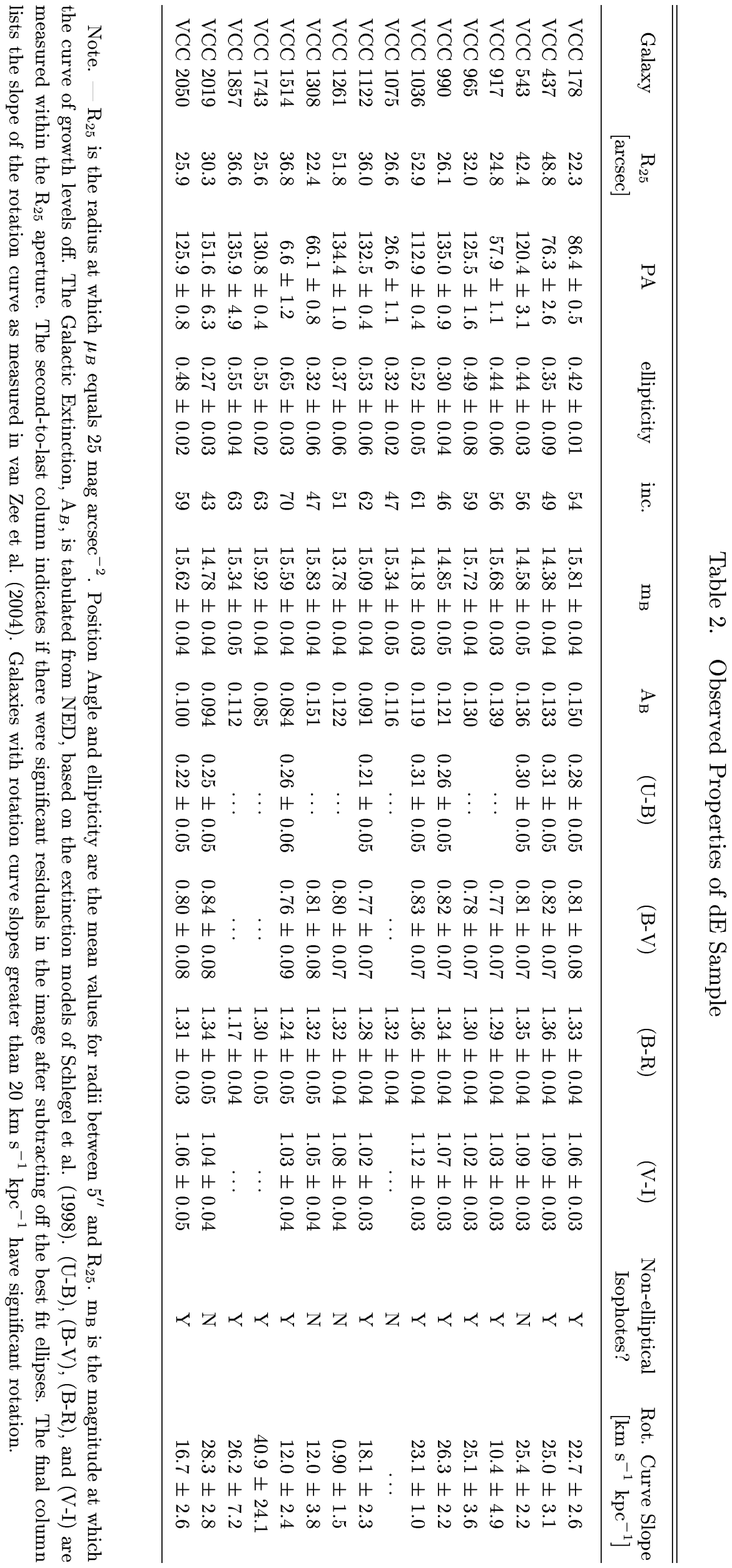,height=25.cm}

\psfig{figure=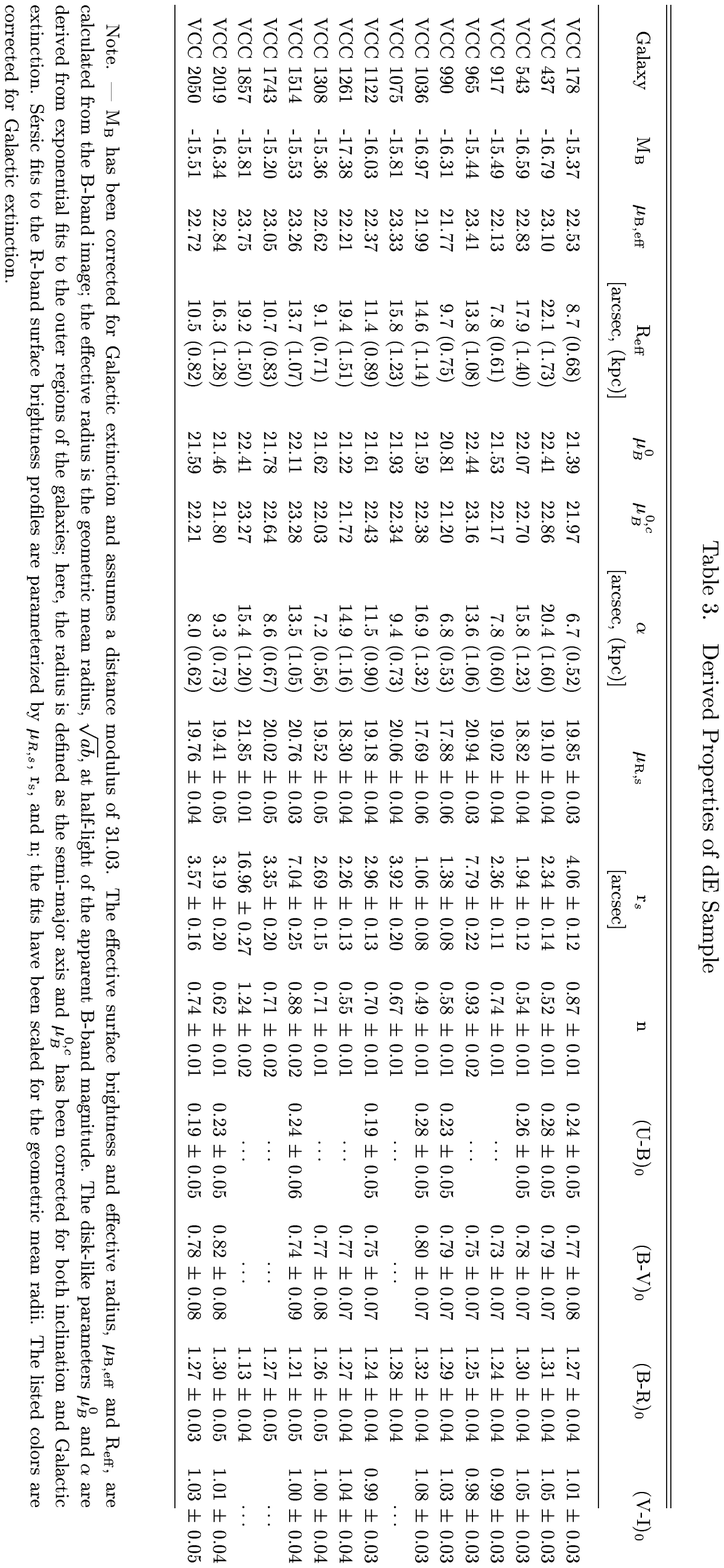,height=25.cm}

\psfig{figure=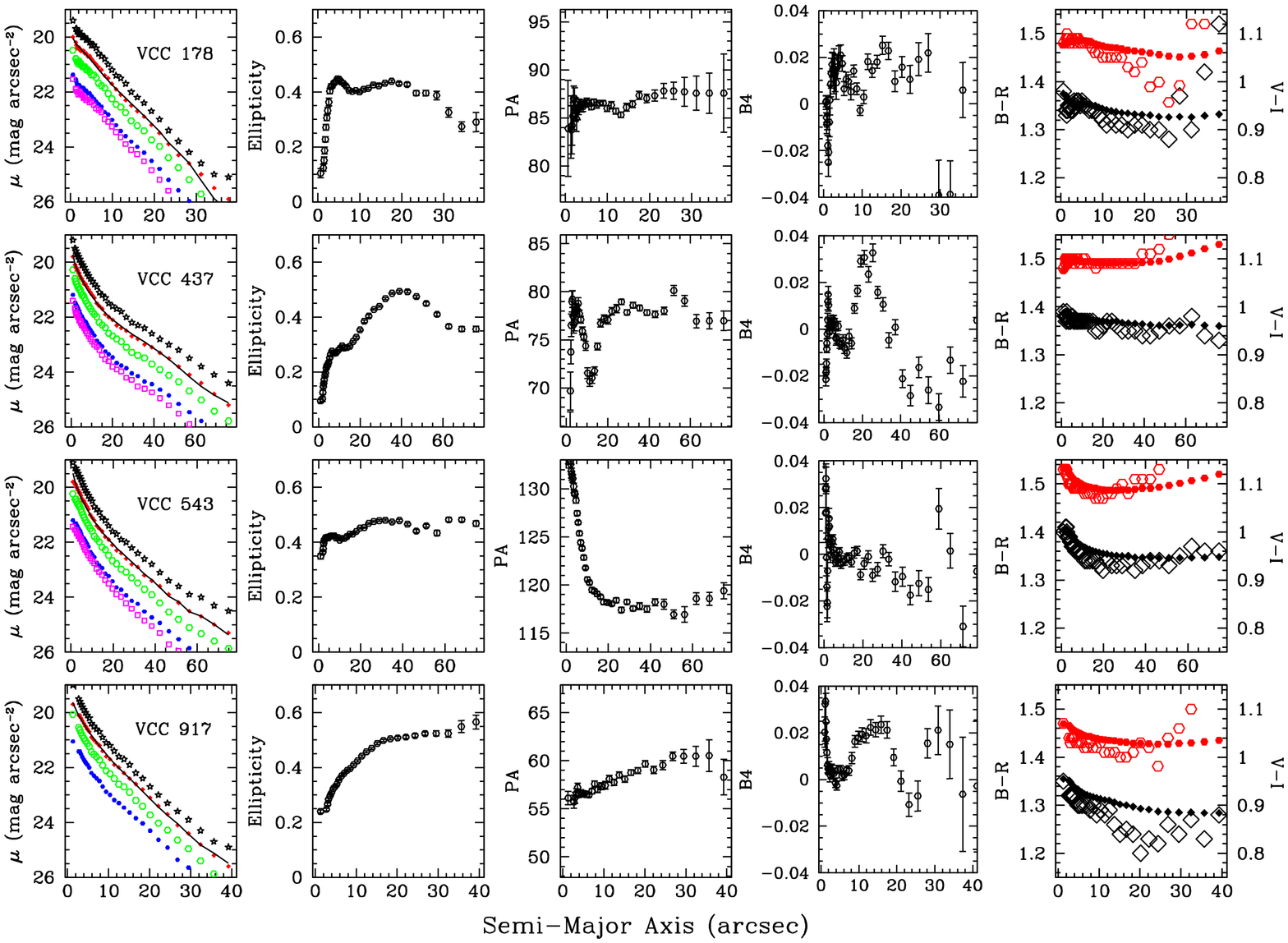,height=13.cm,angle=-90.,bbllx=40pt,bblly=60pt,bburx=580pt,bbury=800pt,clip=t}
%\plotone{vanzee.fig1_1.ps}
\figcaption[vanzee.fig1.ps]{
Surface photometry for 16 dwarf elliptical galaxies.  The left-most panels show the surface
brightness as a function of radius for U (open squares), B (filled hexagons), V (open hexagons),
R (filled diamonds), and I (stars).  The middle panels show the ellipticity, position angle,
and B4 as a function of radius for the ELLIPSE fits to the R-band images.   
The right-most panels show the observed B-R color (black diamonds) and V-I color
(red hexagons) for each elliptical annulus 
(open symbols) and for the integrated light within that
aperture (filled symbols).
\label{fig:surf} }

%\plotone{vanzee.fig1_2.ps}

%\plotone{vanzee.fig1_3.ps}

%\plotone{vanzee.fig1_4.ps}

\psfig{figure=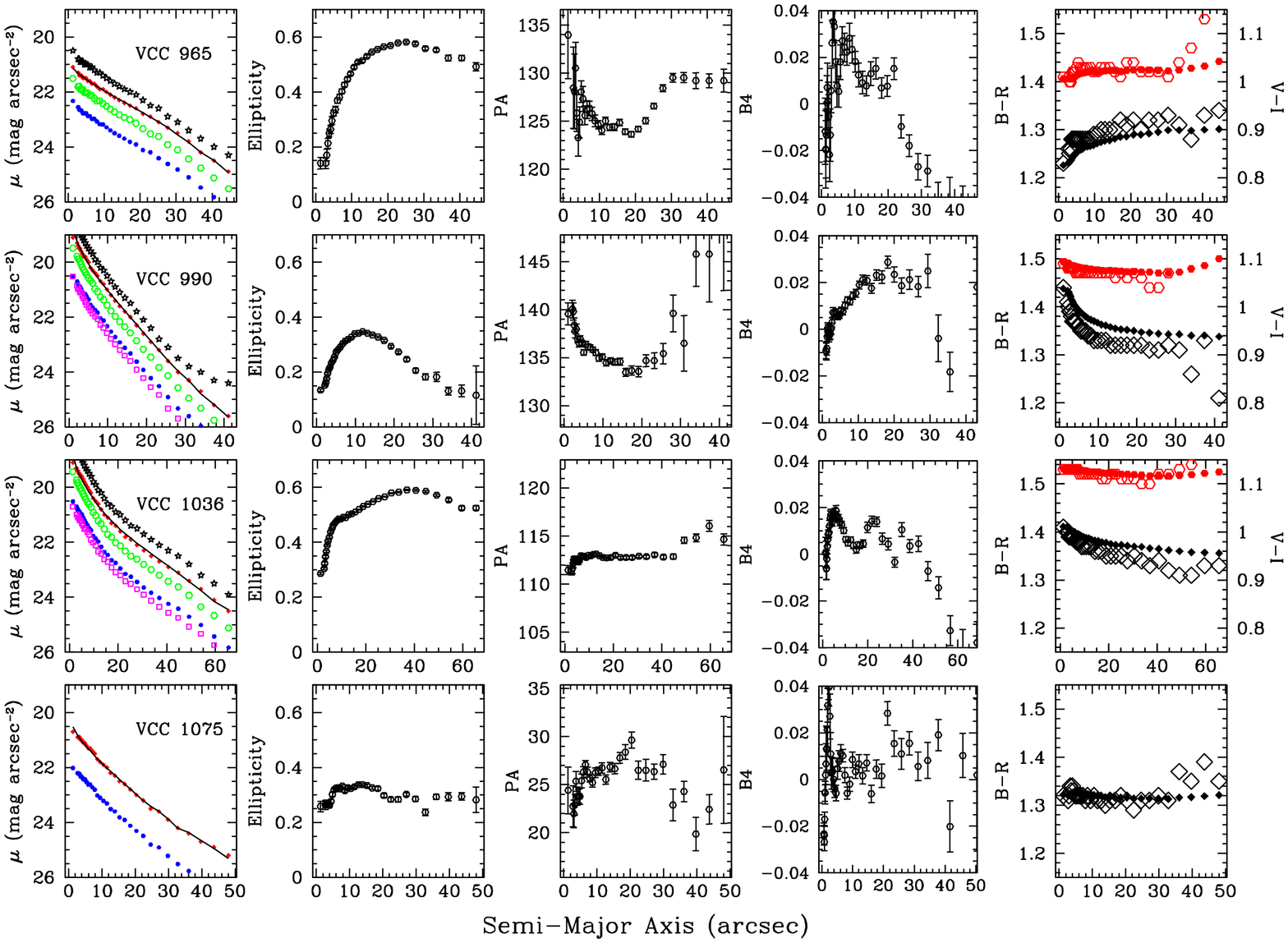,height=13.cm,angle=-90.,bbllx=40pt,bblly=60pt,bburx=580pt,bbury=800pt,clip=t}

\psfig{figure=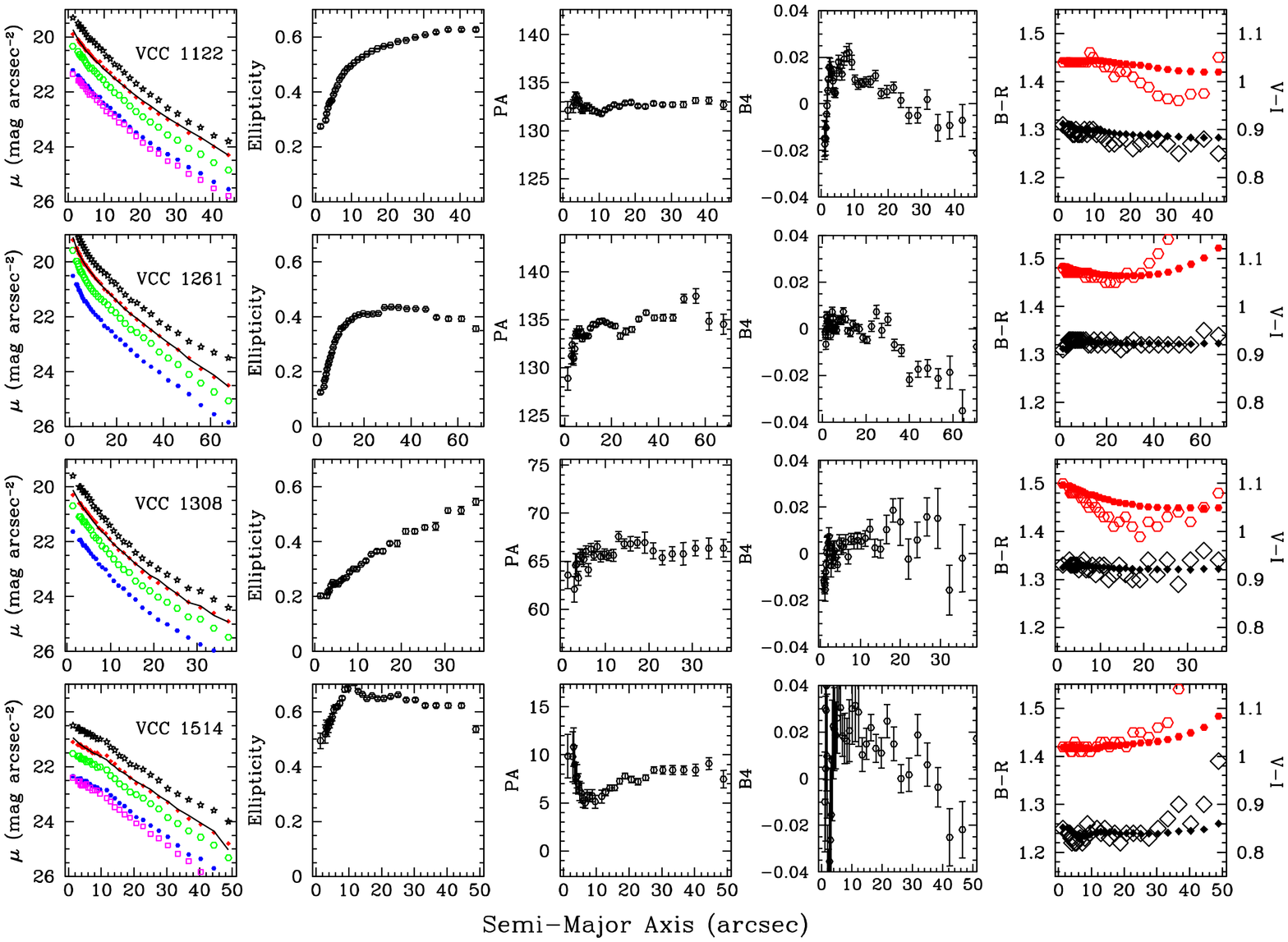,height=13.cm,angle=-90.,bbllx=40pt,bblly=60pt,bburx=580pt,bbury=800pt,clip=t}

\psfig{figure=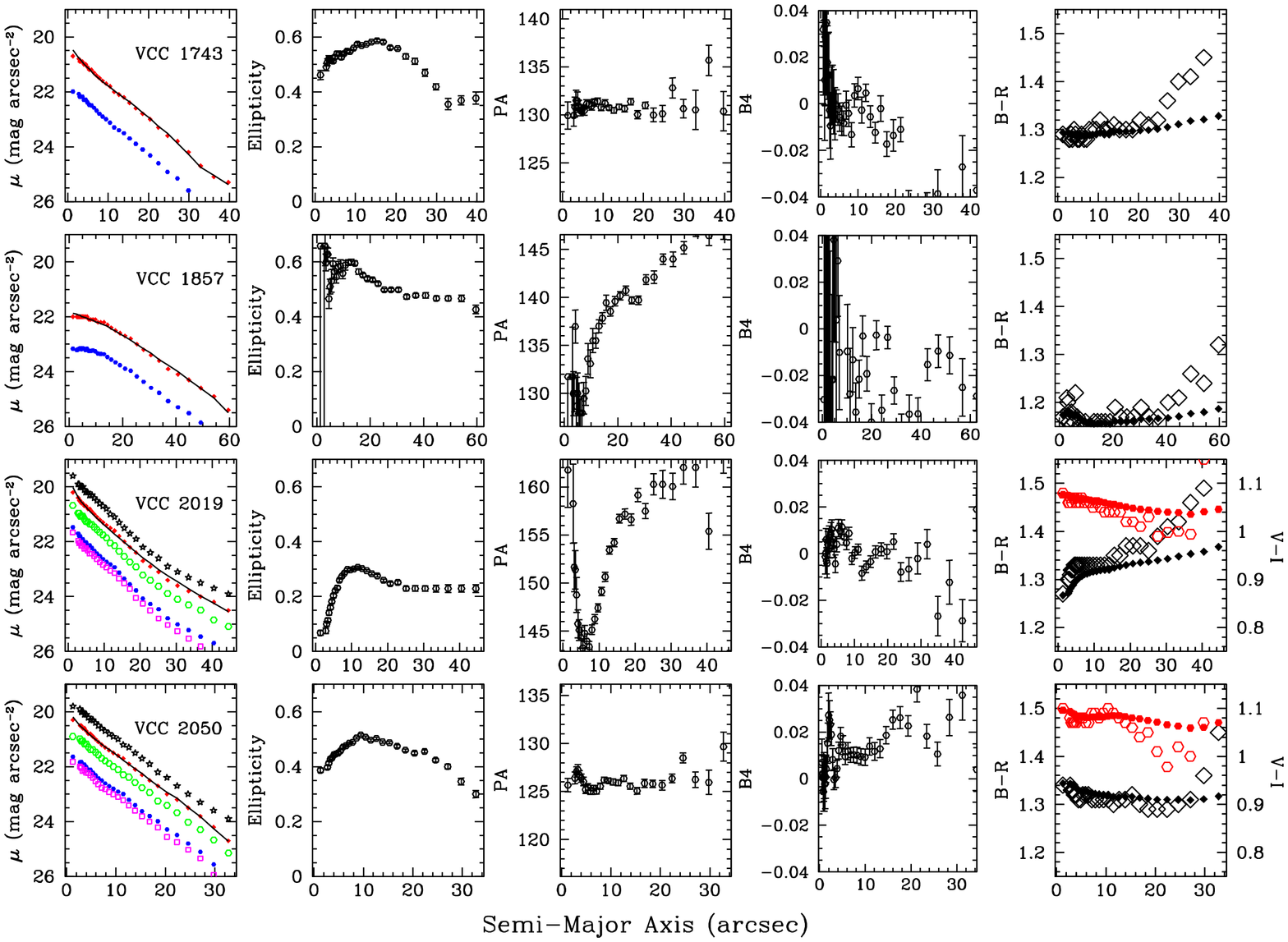,height=13.cm,angle=-90.,bbllx=40pt,bblly=60pt,bburx=580pt,bbury=800pt,clip=t}
                                                                                
\psfig{figure=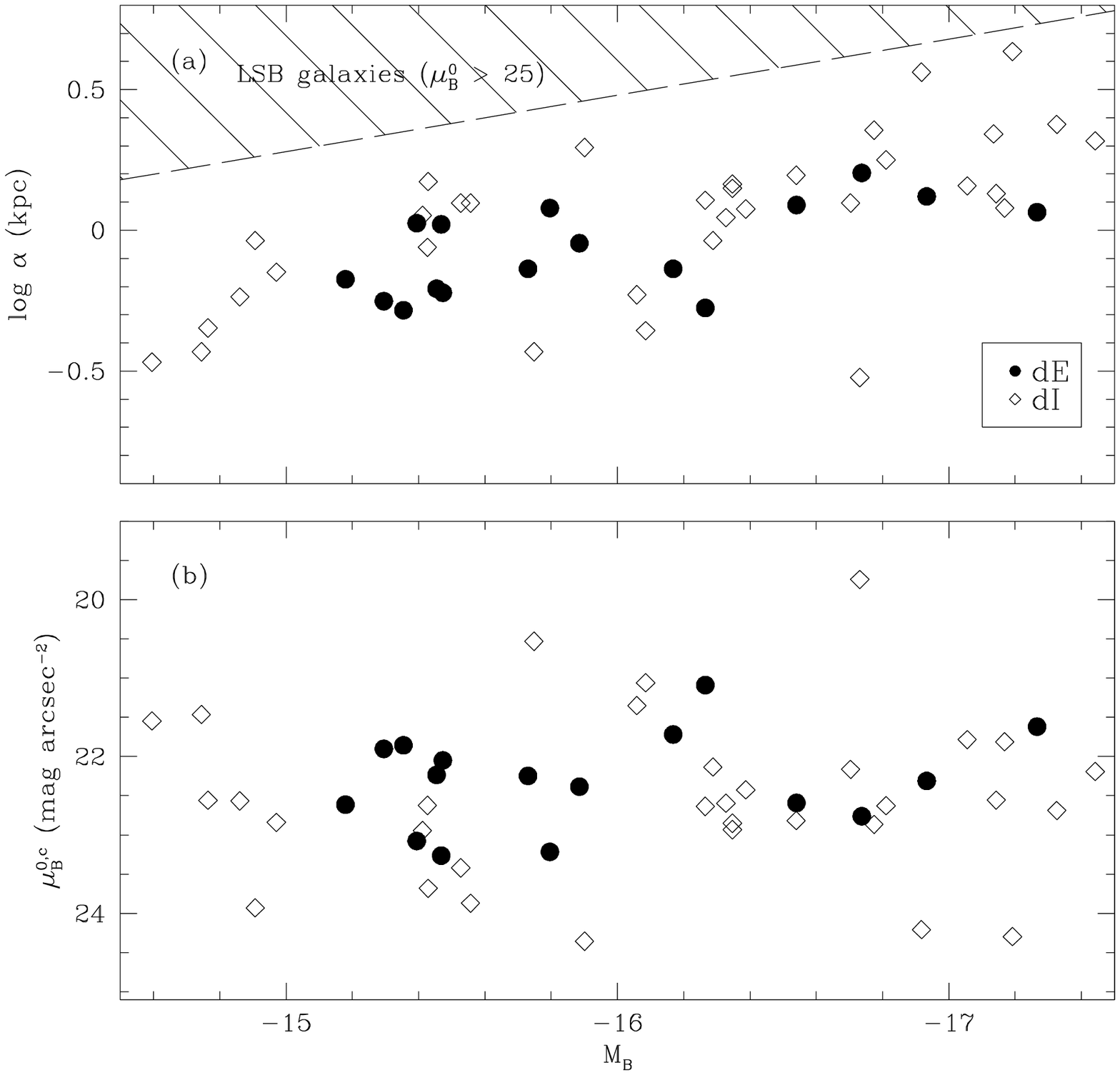,height=16.cm}
\figcaption[vanzee.fig2.ps]{
Structural parameters for exponential fits to dwarf elliptical (filled symbols) 
and dwarf irregular (open symbols; van Zee 2001) galaxies. 
 (a) Scale length as function of absolute magnitude.  (b) Surface brightness as function of absolute
magnitude.  
The dwarf elliptical and dwarf irregular galaxies share a common locus in these diagrams.
\label{fig:struct} }

\psfig{figure=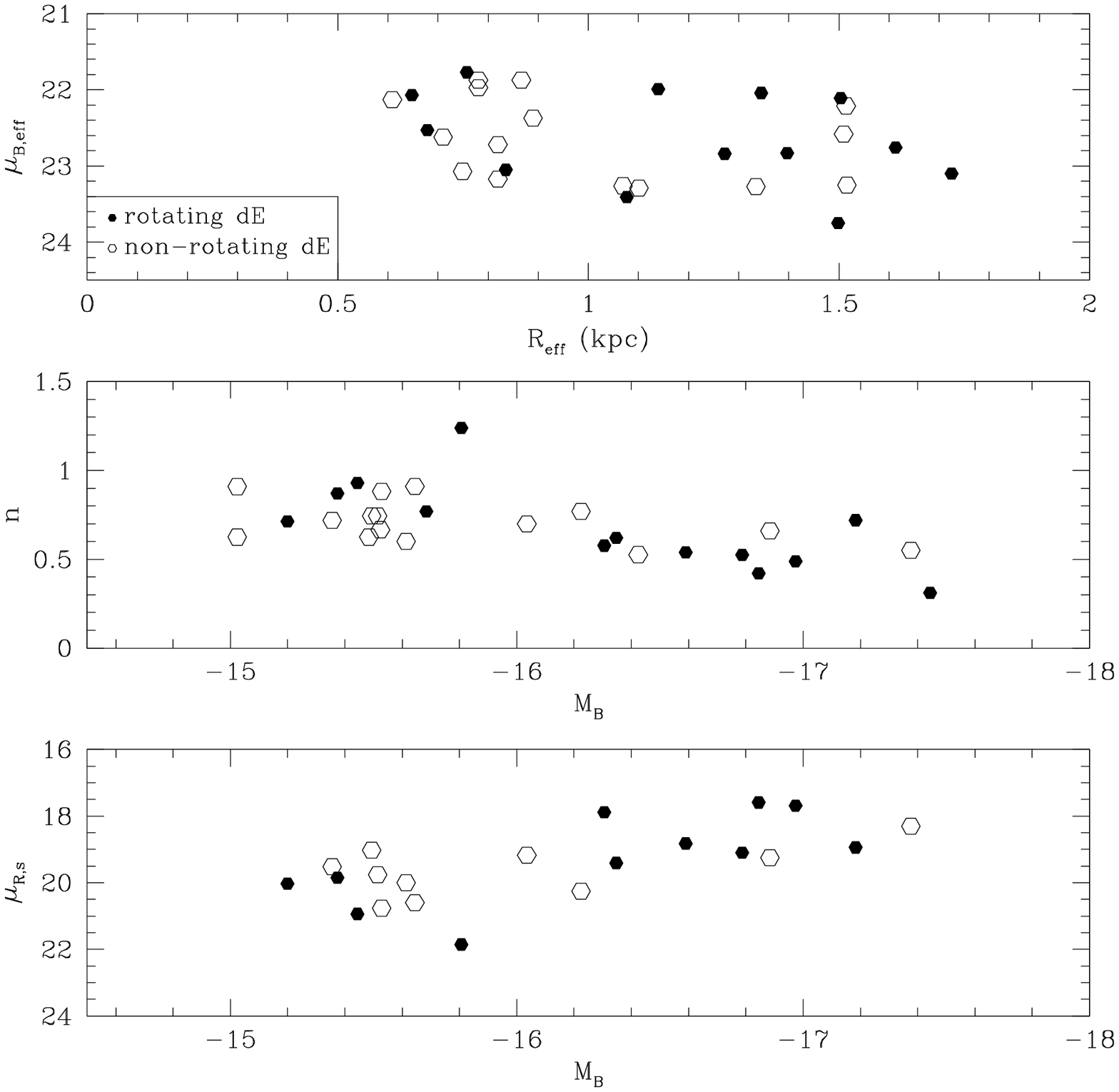,height=16.cm}
\figcaption[vanzee.fig3.ps]{
Structural parameters for S\'ersic fits to dwarf elliptical galaxies in Virgo
with measured kinematic properties (Pedraz et al.\ 2002;
Geha et al.\ 2003; the present sample); rotating (filled hexagons) and non-rotating (open
hexagons) dE galaxies have similar structural properties.  (a) Model independent
parameters of half-light radius and effective surface brightness.  (b) Absolute
magnitude and S\'ersic shape parameter. (c) Absolute magnitude and central surface
brightness from a S\'ersic fit to the R-band surface brightness profile.
The non-rotating and rotating dwarf elliptical galaxies cannot be distinguished
based on structural parameters.
\label{fig:sersic} }

\psfig{figure=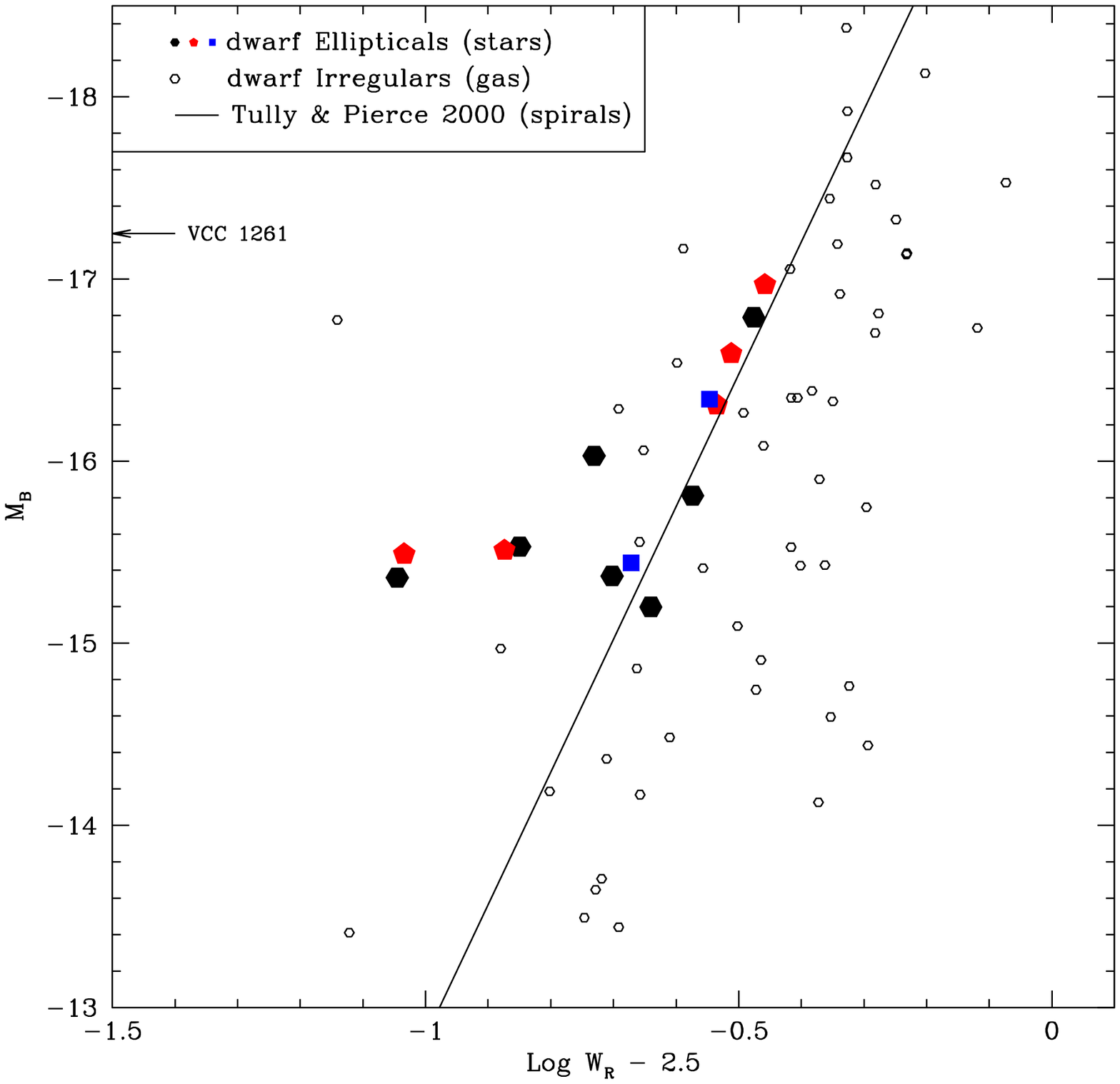,height=16.cm}
\figcaption[vanzee.fig4.ps]{
The luminosity-line width relation for dwarf elliptical galaxies
[note that the values shown are the observed maximum
linewidths, which are possible underestimates of the full rotation velocity;
 see van Zee et al.\ (2004) for a full discussion of possible
correction factors to the observed widths].
Dwarf ellipticals with blue cores (blue squares) and red cores (red pentagons) follow
the same relation as spiral galaxies (line; Tully \& Pierce 2000) and dwarf irregular
galaxies (open dots; van Zee 2001).  
\label{fig:rot}}

\psfig{figure=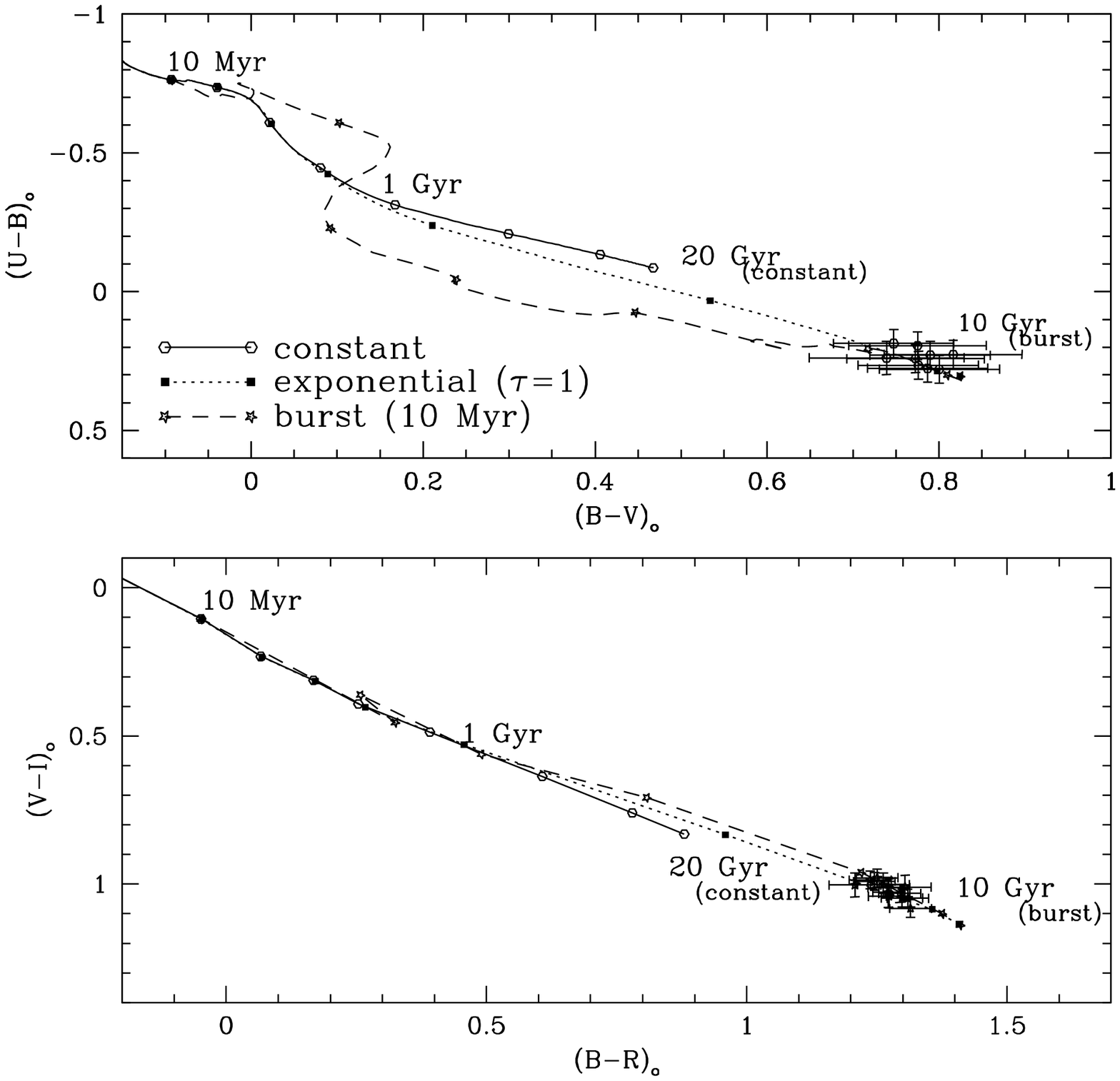,height=15.cm}
\figcaption[vanzee.fig5.ps]{
Optical colors of dwarf elliptical galaxies in the Virgo Cluster.  
Three fiducial evolutionary tracks are shown for 
low metallicity (1/5 solar) composite stellar populations 
from the Bruzual \& Charlot (2003) models. 
The models include a short (10 Myr) burst of star formation (long dashed lines), 
an exponentially decreasing star formation rate, with and e--folding time of 1 Gyr 
(short dashed lines), and a constant star formation rate (solid lines).  
The models are marked every 0.5 dex, and time increases from the upper left 
hand corner to the lower right hand corner.  The short burst and exponentially 
decreasing star formation rates both result in a red
galaxy after a Hubble time, while a constant star formation rate results in a blue
system.  The three star formation histories can be separated in the UBV color-color
diagram but all three models are degenerate in the color-color space of (B-R) vs. (V-I).
The observed colors are consistent with an evolved stellar population with
an age of approximately 6 Gyr.
\label{fig:evol} }

\psfig{figure=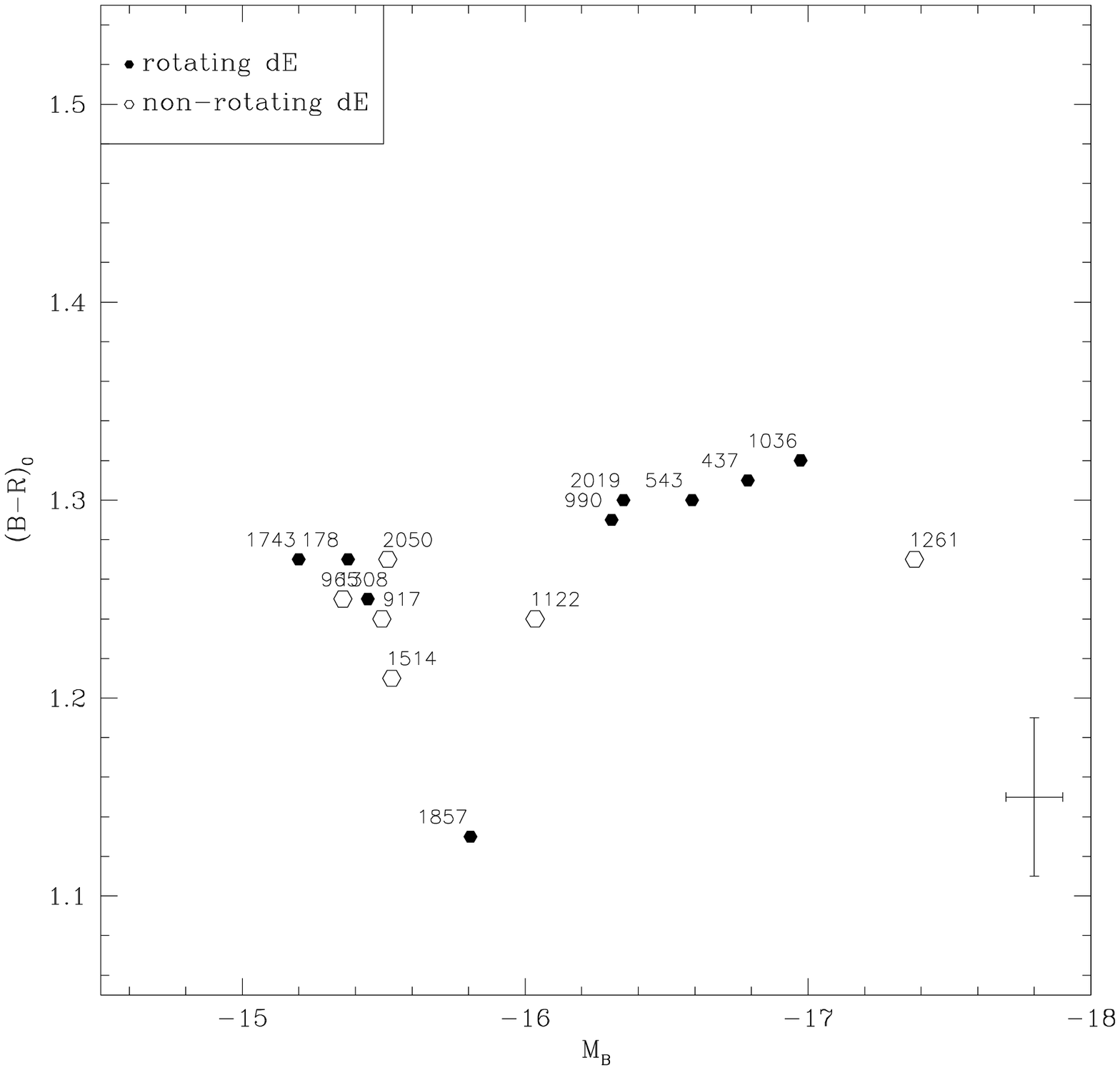,height=16.cm}
\figcaption[vanzee.fig6.ps]{
Color-magnitude diagram for dwarf elliptical galaxies with measured kinematic
properties.
The more luminous dEs are slightly redder than the lower luminosity
galaxies, indicating a change in age or metallicity of the dominant
stellar population.  The two significant
outliers in this diagram are VCC 1857 and VCC 1261, one of which
appears to be rotationally supported while the other has no
evidence of rotation (kinematics from major-axis spectroscopy only).  These observations
indicate that there are no clear differences in the optical colors 
of rotating (filled hexagons) and non--rotating (open hexagons) 
dwarf elliptical galaxies.
\label{fig:colmag}}

\psfig{figure=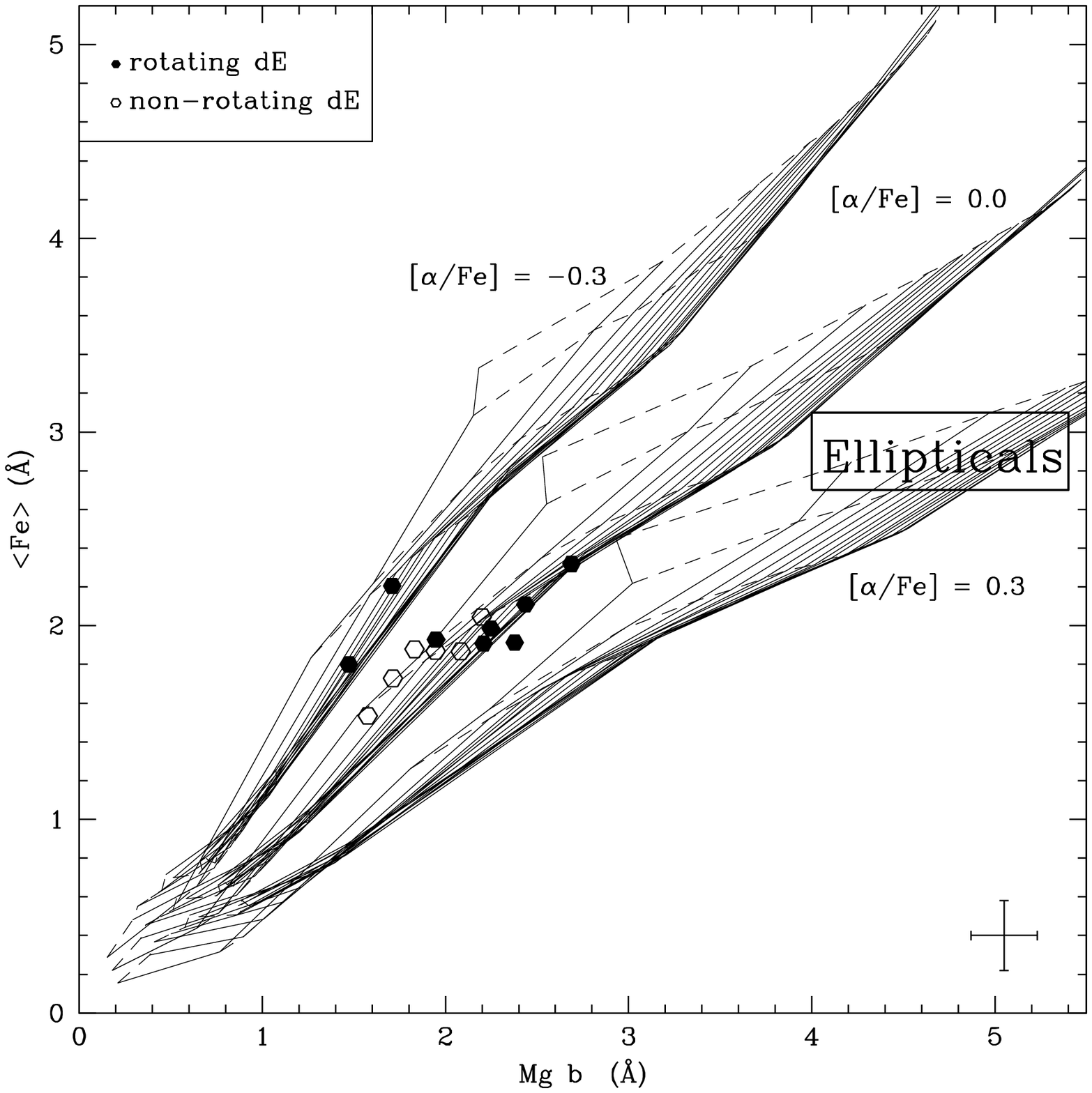,height=16.cm}
\figcaption[vanzee.fig7.ps]{
Mgb and $<Fe>$ Lick/IDS indices for Virgo dwarf elliptical galaxies.
The locus for giant elliptical galaxies from Trager et al. (2000)
is also shown.
The stellar population models of Thomas et al.\ (2003b) are shown
for several ages (t = 1 to 15 Gyr in 1 Gyr increments) and for several
[$\alpha$/Fe] models ([$\alpha$/Fe] = -0.3, 0.0, and 0.3 dex).  The
majority of dwarf elliptical galaxies are well fit by models
with solar [$\alpha$/Fe] ratios.
\label{fig:alpha}}

\psfig{figure=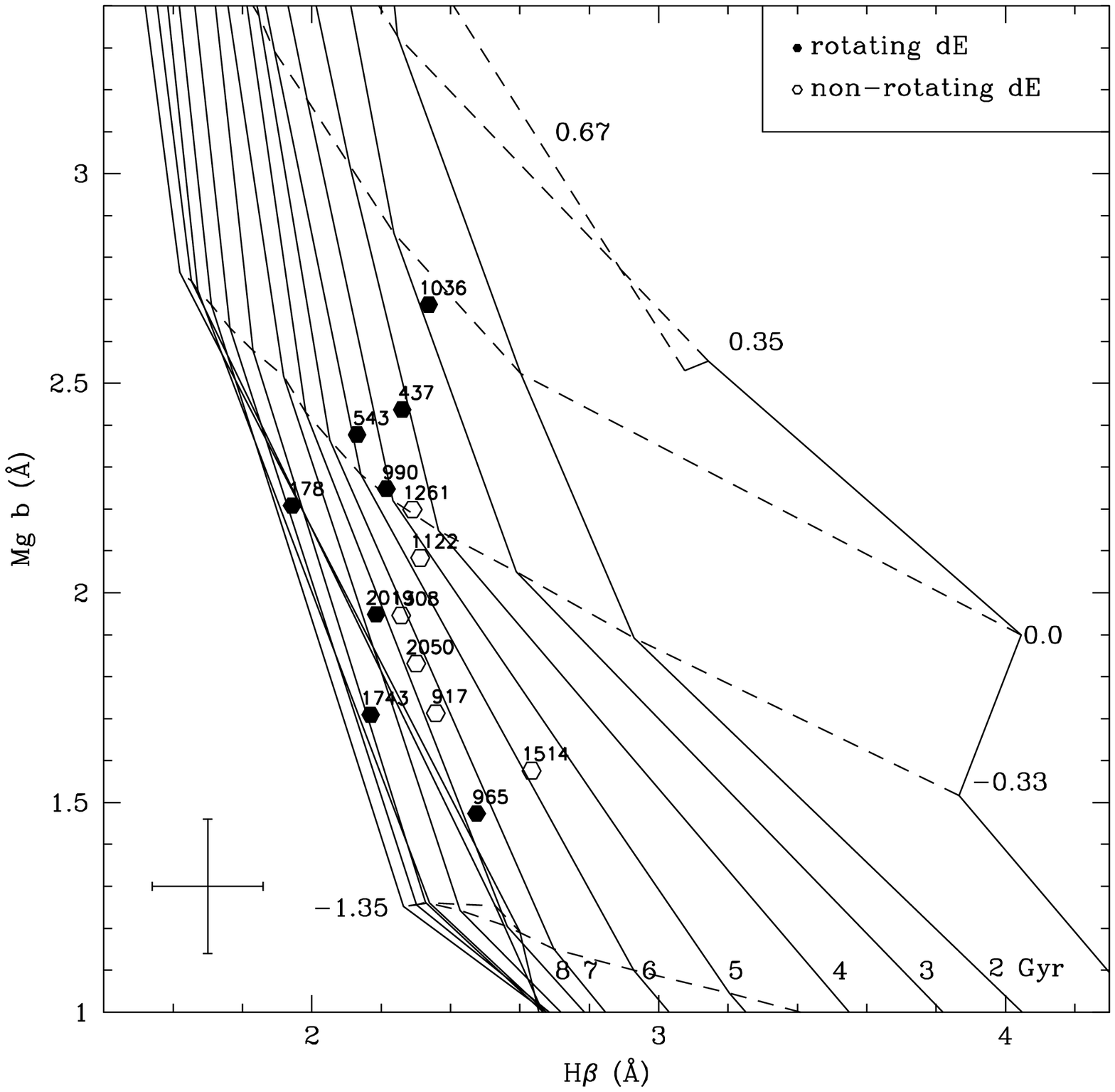,height=16.cm}
\figcaption[vanzee.fig8.ps]{
Mgb and H$\beta$ Lick/IDS indices for Virgo dwarf elliptical galaxies.
Stellar population models with solar [$\alpha$/Fe] ratios (Thomas et al.\ 2003b) 
are shown for several metallicities ([Fe/H] = -0.33, 0.0, 0.33, and 0.67 dex) and
for several ages (t = 1 to 15 Gyr in 1 Gyr increments).
The majority of dwarf elliptical galaxies contain low metallicity ([Fe/H] $<$ -0.33)
evolved stellar populations (t $\sim$ 5-7 Gyr).
\label{fig:metals}}

\end{document}